\documentclass[%
 aip,
pof,
 amsmath,amssymb,
preprint,%
]{revtex4-1}

\usepackage {CJK}

\usepackage{graphicx}% Include figure files
\usepackage{dcolumn}% Align table columns on decimal point
\usepackage{bm}% bold math
\usepackage[utf8]{inputenc}
\usepackage[T1]{fontenc}
\usepackage{mathptmx}

\usepackage{bmpsize}

\begin{document}

\begin{CJK*}{UTF8}{} % Use default fonts from CJK (see below)

\preprint{Preprint submit to \emph{Physics of Fluids}}

\title{Production and transport of vorticity in two-dimensional Rayleigh-B\'enard convection cell}
% Force line breaks with \\

\author{Ao Xu}
 \affiliation{School of Aeronautics, Northwestern Polytechnical University, Xi'an 710072, China}%
  \affiliation{Institute of Extreme Mechanics, Northwestern Polytechnical University, Xi'an 710072, China}%

\author{Ben-Rui Xu}
 \affiliation{School of Aeronautics, Northwestern Polytechnical University, Xi'an 710072, China}%

\author{Li-Sheng Jiang}
 \affiliation{School of Aeronautics, Northwestern Polytechnical University, Xi'an 710072, China}%

\author{Heng-Dong Xi}
\email{Corresponding author: hengdongxi@nwpu.edu.cn (Heng-Dong Xi)}
 \affiliation{School of Aeronautics, Northwestern Polytechnical University, Xi'an 710072, China}%
 \affiliation{Institute of Extreme Mechanics, Northwestern Polytechnical University, Xi'an 710072, China}%

\date{\today}% It is always \today, today,

\begin{abstract}
We present a numerical study of vorticity production and transport in the two-dimensional Rayleigh-B\'enard (RB) convection.
Direct numerical simulations are carried out in the Rayleigh number ($Ra$) range $10^{5}\le Ra \le 10^{6}$, the Prandtl number ($Pr$) of 0.71, and the aspect ratio ($\Gamma$) of the convection cell range $0.75\le \Gamma \le 6$.
We found that the flow structure and temperature distribution vary with $\Gamma$ greatly due to multiple vortices interaction.
Further investigation on the vorticity production and transport reveals that, in the RB convection, in addition to the vorticity production due to wall shear stress, buoyancy produces significant vorticity in the bulk region.
The produced vorticity is transported via advection and diffusion.
An interesting finding is that the main vortices and the corner vortices can be visualized via the contour of buoyancy-produced vorticity.
Although a vigorous definition of the vortex is still lacking in the community, our efficient vortex visualization approach in the RB convection may shed light on further research toward vortex identification.
We also found that the spatial distribution of vorticity flux along the wall is positively correlated with that of the Nusselt number ($Nu$), suggesting the amount of vorticity that enters the flow is directly related to the amount of thermal energy that enters the flow.
\footnote{
%%The following article has been submitted to Physics of Fluids.
%%After it is published, it will be found at Link (https://publishing.aip.org/resources/librarians/products/journals/).
This article may be downloaded for personal use only.
Any other use requires prior permission of the author and AIP Publishing.
This article appeared in Xu et al., Phys. Fluids \textbf{34}, 013609 (2022) and may be found at \url{https://doi.org/10.1063/5.0072873}.
}
\end{abstract}

\maketitle
\end{CJK*}

\section{\label{Section1}Introduction}

Thermal convection occurs ubiquitously in nature and has wide applications in industry, such as convection over the city, \cite{omidvar2020plume} convection in the indoor environment, \cite{bhagat2020effects,xu2020transport} convection in the heat exchanger reactor, \cite{shah2003fundamentals} and so on.
A canonic flow systems for studying thermal convection is Rayleigh-B\'enard (RB) convection, \cite{lohse2010small,xia2013current} in which the fluid is heated from the bottom wall and cooled from the top wall.
In thermal convection cells, approaches to enhance heat transfer efficiency include creating roughness on the heating walls, \cite{zhang2018surface,dong2020influence} geometrical confinement of the convection cell, \cite{huang2013confinement,huang2016effects} vibration-induced boundary-layer destabilization, \cite{wang2020vibration,wu2021phase} and many others.

Previous studies have revealed the connections between heat transfer efficiency (in terms of dimensionless Nusselt number, $Nu$) and flow structures. \cite{sun2005azimuthal,xi2008flow,xi2016higher,xu2020correlation,van2011connecting,van2012flow,zou2019boundary,zhu2021flow}
For example, Sun \emph{et al.} \cite{sun2005azimuthal} measured the $Nu$ in both leveled and tilted cylindrical cells, in which the large-scale circulation is either azimuthal or locked in a particular orientation, respectively.
They found that the $Nu$ is larger in the leveled cell than that in the tilted one, thus demonstrating that different flow structures can give rise to different values of $Nu$.
Xi \emph{et al.} \cite{xi2008flow} further observed that both the single-roll and the double-roll flow structures exist in the large-scale flow.
They examined the averaged $Nu$ corresponding to a particular flow structure, and they found that the single-roll flow structure is more efficient for heat transfer than the double-roll structure.
Later, Xi \emph{et al.} \cite{xi2016higher} found that $Nu$ has a momentary overshoot above its average value during flow reversal event, which is also numerically verified by Xu \emph{et al.}  \cite{xu2020correlation}
The overshoot in $Nu$ was attributed to more coherent flow or plumes for the short period of time during reversal.
van der Poel \emph{et al.} \cite{van2011connecting,van2012flow} simulated the aspect ratio dependence of heat transfer efficiency in a two-dimensional (2D) square cell.
They conditionally averaged $Nu$ based on flow structures, and they found that heat transfer is more efficient with less vertically arranged vortex or less horizontally elongated vortex.
Xu \emph{et al.} \cite{xu2020correlation} adopted the Fourier mode decomposition and the proper orthogonal decomposition to extract the coherent flow structure.
They showed that the single-roll mode, the horizontally stacked double-roll mode, and the quadrupolar flow mode are more efficient for heat transfer on average; in contrast, the vertically stacked double-roll mode is inefficient for heat transfer on average.  \cite{xu2020correlation}

On the other hand, it is advantageous to interpret the fluid flows in terms of vorticity, which is defined as the curl of velocity $\omega=\nabla \times \mathbf{u}$.  \cite{panton2013incompressible,davidson2015turbulence}
The vorticity represents the rotation of a fluid particle, and only the shear stresses can rotate the fluid particle.
As for the pressure and the normal viscous stresses, they act through the center of the fluid particle and cannot rotate fluid particles.
Thus, analyzing vorticity dynamics will reveal physical mechanisms for fluid flows and associate transport processes (e.g., heat transfer, mass transfer, ion transfer) that may be hidden from the velocity and pressure fields.
In the industry, various vortex generators have been designed to enhance transfer in heat exchanger reactors;  \cite{lemenand2018vorticity,karkaba2020multi} however, most of those studies focused on a global relationship between the velocity field and heat transfer efficiency, and the role of vorticity transport in heat transfer enhancement remains unclear.

In this work, we will provide a comprehensive analysis of vorticity production and transport in the simple yet canonic RB convection.
Our motivation is to reveal the connections between heat transfer efficiency and the vorticity field.
The rest of the paper is organized as follows.
In Sec. \ref{Section2}, we will present the numerical method for simulating incompressible thermal convection.
The in-house numerical solver based on the lattice Boltzmann (LB) method will be introduced.
In Sec. \ref{Section3}, we will first provide general flow and heat transfer features in the convection cells; after that, we will analyze vorticity production due to wall shear stress and buoyancy, as well as vorticity transport due to advection and diffusion.
In Sec. \ref{Section4}, the main findings of the present work are summarized.

\section{\label{Section2}Numerical method}

\subsection{Mathematical model for thermal convection}

We consider incompressible thermal flows under the Boussinesq approximation. The temperature is treated as an active scalar and its influence on the velocity field is realized through the buoyancy term. The viscous heat dissipation and compression work are neglected, and all the transport coefficients are assumed to be constants. Then, the governing equations can be written as
\begin{subequations}
\begin{align}
& \nabla \cdot \mathbf{u}=0 \\
& \frac{\partial \mathbf{u}}{\partial t}+\mathbf{u}\cdot \nabla \mathbf{u}=-\frac{1}{\rho_{0}}\nabla P+\nu \nabla^{2}\mathbf{u}+g\beta_{T}(T-T_{0})\hat{\mathbf{y}} \\
& \frac{\partial T}{\partial t}+\mathbf{u}\cdot \nabla T=\alpha_{T} \nabla^{2} T
\end{align} \label{Eq.NS}
\end{subequations}
where $\mathbf{u}=(u,v)$, $P$, and $T$ are velocity, pressure, and temperature of the fluid, respectively.
$P_{0}$, $\rho_{0}$, and $T_{0}$ are the reference pressure, density and temperature, respectively.
$\beta_{T}$, $\nu$, and $\alpha_{T}$ are the thermal expansion coefficient, kinematic viscosity, and thermal diffusivity, respectively.
$g$ is the gravity acceleration value.
$\hat{\mathbf{y}}$ is the unit vector parallel to the gravity.
With the following non-dimensional group
\begin{equation}
    \begin{split}
& \mathbf{x}^{*} = \mathbf{x}/H, \ \ \ t^{*}=t/\sqrt{H/(g\beta_{T}\Delta_{T})}, \ \ \ \mathbf{u}^{*}=\mathbf{u}/\sqrt{g\beta_{T}\Delta_{T}H}, \\
& P^{*}=(P-P_{0})/(\rho_{0}g\beta_{T}\Delta_{T}H), \ \ \ T^{*}=(T-T_{0})/\Delta_{T} \\
    \end{split}
\label{Eq.non-dimensional}
\end{equation}
Then, Eq. (\ref{Eq.NS}) can be rewritten in dimensionless form as
\begin{subequations}
\begin{align}
& \nabla \cdot \mathbf{u}^{*}=0 \\
& \frac{\partial \mathbf{u}^{*}}{\partial t^{*}}+\mathbf{u}^{*}\cdot \nabla \mathbf{u}^{*}
=-\nabla P^{*}+\sqrt{\frac{Pr}{Ra}} \nabla^{2}\mathbf{u}^{*}+T^{*}\hat{\mathbf{y}} \\
& \frac{\partial T^{*}}{\partial t^{*}}+\mathbf{u}^{*}\cdot \nabla T^{*}=\sqrt{\frac{1}{PrRa}} \nabla^{2} T^{*}
\end{align} \label{Eq.NS-dimensionless}
\end{subequations}
Here, $H$ is the cell height and $\Delta_{T}$ is the temperature difference between heating and cooling walls.
$P_{0}$ is the reference pressure.
In this paper, unless otherwise stated, the dimensionless variable is denoted with a superscript star.
The two dimensionless parameters are the Rayleigh number ($Ra$) and the Prandtl number ($Pr$), which are defined as
\begin{equation}
Ra = \frac{g\beta_{T}\Delta_{T}H^{3}}{\nu \alpha_{T}}, \ \ \ Pr=\frac{\nu}{\alpha_{T}}
\end{equation}

\subsection{The lattice Boltzmann method for incompressible thermal flows}
The LB method to solve fluid flows and heat transfer is based on the double distribution function approach.
The advantages of the LB method include easy implementation and parallelization as well as low numerical dissipation. \cite{chen1998lattice,aidun2010lattice,xu2017lattice}
Specifically, we chose a D2Q9 model for the Navier-Stokes equations to simulate fluid flows and a D2Q5 model for the energy equation to simulate heat transfer.
To enhance the numerical stability, the multi-relaxation-time collision operator is adopted in the evolution equations of both density and temperature distribution functions.
The evolution equation of the density distribution function is written as
\begin{equation}
  f_{i}(\mathbf{x}+\mathbf{e}_{i}\delta_{t},t+\delta_{t})-f_{i}(\mathbf{x},t)=-(\mathbf{M}^{-1}\mathbf{S})_{ij}\left[\mathbf{m}_{j}(\mathbf{x},t)-\mathbf{m}_{j}^{(\text{eq})}(\mathbf{x},t)\right]
  +\delta_{t}F_{i}^{'} \label{Eq.MRT}
\end{equation}
where $f_{i}$ is the density distribution function.
$\mathbf{x}$ is the fluid parcel position, $t$ is the time, and $\delta_{t}$ is the time step.
$\mathbf{e}_{i}$ is the discrete velocity along the $i$th direction.
$\mathbf{M}$ is a $9\times 9$ orthogonal transformation matrix that projects the density distribution function $f_{i}$ and its equilibrium $f_{i}^{(\text{eq})}$ from the velocity space onto the moment space, such that $\mathbf{m}=\mathbf{M}\mathbf{f}$ and $\mathbf{m}^{(\text{eq})}=\mathbf{M}\mathbf{f}^{(\text{eq})}$.
$\mathbf{S}$ is the diagonal relaxation matrix.
The macroscopic density $\rho$ and velocity $\mathbf{u}$ are obtained from the density distribution function as
\begin{equation}
\rho=\sum_{i=0}^{8}f_{i}, \ \ \mathbf{u}=\frac{1}{\rho}\left( \sum_{i=0}^{8}\mathbf{e}_{i}f_{i}+\mathbf{F}/2 \right)
\end{equation}
The evolution equation of temperature distribution function is written as
\begin{equation}
  g_{i}(\mathbf{x}+\mathbf{e}_{i}\delta_{t},t+\delta_{t})-g_{i}(\mathbf{x},t)=-(\mathbf{N}^{-1}\mathbf{Q})_{ij}\left[\mathbf{n}_{j}(\mathbf{x},t)-\mathbf{n}_{j}^{(\text{eq})}(\mathbf{x},t)\right]
  \label{Eq.MRT_T}
\end{equation}
where $g_{i}$ is the temperature distribution function.
$\mathbf{N}$ is a $5\times 5$ orthogonal transformation matrix that projects the temperature distribution function $g_{i}$ and its equilibrium $g_{i}^{(\text{eq})}$ from the velocity space onto the moment space, such that $\mathbf{n}=\mathbf{N}\mathbf{g}$ and $\mathbf{n}^{(\text{eq})}=\mathbf{N}\mathbf{g}^{(\text{eq})}$.
$\mathbf{Q}$ is the diagonal relaxation matrix.
The macroscopic temperature $T$ is obtained from the temperature distribution function as
\begin{equation}
T=\sum_{i=0}^{4}g_{i}
\end{equation}
Our in-house LB solver is accelerated with OpenACC directives to utilize the computing power of GPUs.  \cite{xu2017accelerated}
More numerical details of the LB method and validation of the in-house solver can be found in our previous work.  \cite{xu2019lattice,xu2019statistics,xu2021tristable}

\subsection{Simulation settings}
The top and bottom walls of the convection cell are kept at constant cold and hot temperature, respectively, while the other two vertical walls are adiabatic.
All four walls impose no-slip velocity boundary conditions.
The dimension of the cell is $L \times H$.
Here, $L$ is the cell length.
Simulation results are provided for the $Ra$ in the range  $10^{5} \le Ra \le 10^{6}$, the $Pr$ of 0.71, and cell aspect ratio ($\Gamma=L/H$) in the range  $0.75 \le \Gamma \le 6$.
The criterion for reaching a steady state is
\begin{equation}
\frac{\sum_{i}\parallel \mathbf{u}(x_{i},t+2000\delta_{t})-\mathbf{u}(x_{i},t) \parallel_{2}}{\sum_{i}\parallel \mathbf{u}(x_{i},t) \parallel_{2}} < 10^{-9}, \ \ \
\frac{\sum_{i}| T(x_{i},t+2000\delta_{t})-T(x_{i},t)|_{2}}{\sum_{i}| T(x_{i},t) |_{2}} < 10^{-9}
\end{equation}
For unsteady flows, after the flows reach the statistically stationary state, we take average time of $t_{avg}$ to obtain statistically convergent results.
The time-averaged value of variable $\phi$ is define as $\bar{\phi}=\int_{t_{1}}^{t_{2}}\phi dt / (t_{2}-t_{1})$, where $t_{2}-t_{1}=t_{avg}$ is the time duration of the flow in the fully developed state.

We first refine the grid to check the convergence behaviors of the in-house LB solvers.
We calculated response parameters including the Nusselt number ($Nu$) and the Reynolds number ($Re$), which describe the heat transfer efficiency and flow strength, respectively,
\begin{equation}
Nu=\sqrt{Ra \cdot Pr} \langle v^{*}T^{*} \rangle_{V,t}+1, \ \ \
Re=\frac{\sqrt{\langle (u^{2}+v^{2}) \rangle_{V,t}}H}{\nu}
\end{equation}
Here, $\langle \cdot \rangle_{V,t}$ denotes the spatial and temporal average.
In Table \ref{tab:converge}, we list the steady solution of $Nu$, $Re$, and the absolute value of vorticity at cell center ($|\omega_{c}|$) obtained at $Ra=10^{6}$, $Pr=0.71$, and $\Gamma=1$.
From Table \ref{tab:converge}, we can see that results calculated from the LB solver monotonous converge.

\begin{table}[htbp]
  \centering
  \caption{Spatial convergence of the in-house lattice Boltzmann (LB) solver. The columns include grid number, the Nusselt number ($Nu$), the Reynolds number ($Re$), and the absolute value of vorticity at cell center ($|\omega_{c}|$) for $Ra=10^{6}$, $Pr=0.71$, and $\Gamma=1$.}
    \begin{tabular}{cccc}
\hline
    Grid  & \ \ $Nu$ \ \ & \ \ $Re$ \ \ & $ \ \ |\omega_{c}|$ \ \ \\
\hline
    $32^{2}$ \ \   &  \ \ 6.4393 \ \ &  \ \ 274.40 &  \ \ 1.5280 \ \ \\
    $64^{2}$ \ \   &  \ \ 6.3489 \ \ &  \ \ 275.36 &  \ \ 1.5114 \ \ \\
    $128^{2}$ \ \  &  \ \ 6.3336 \ \ &  \ \ 275.62 &  \ \ 1.5074 \ \ \\
    $256^{2}$ \ \  &  \ \ 6.3299 \ \ &  \ \ 275.70 &  \ \ 1.5066 \ \ \\
    $512^{2}$ \ \  &  \ \ 6.3290 \ \ &  \ \ 275.72 &  \ \ 1.5064 \ \ \\
    $1024^{2}$ \ \ &  \ \ 6.3287 \ \ &  \ \ 275.72 &  \ \ 1.5064 \ \ \\
    $2048^{2}$ \ \ &  \ \ 6.3287 \ \ &  \ \ 275.72 &  \ \ 1.5064 \ \ \\
\hline
    \end{tabular}%
  \label{tab:converge}%
\end{table}%

\begin{figure}[htbp]
\centering
\includegraphics[width=7cm]{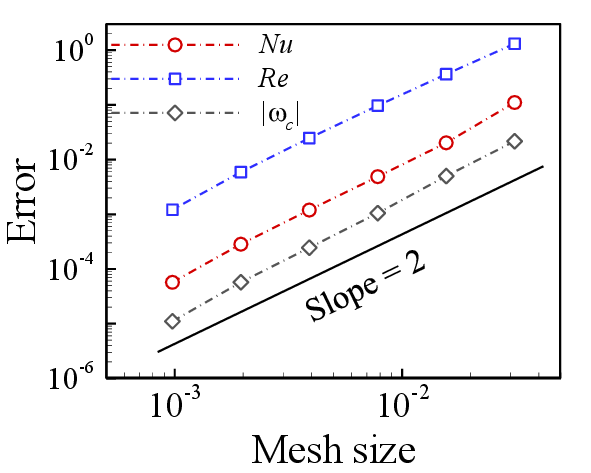}
\caption{\label{Fig-converge-Nu-Re} Spatial convergence of the in-house LB solver. The solid line is to guide the eye.}
\end{figure}

To quantitatively evaluate the spatial convergence order, we assume the results obtained at the largest grid of $2048^{2}$ are accurate and calculate the error as  $\text{Error}(\delta x)=|\mathcal{F}(\delta x)-\mathcal{F}(1/2048)|$.
Here, $\mathcal{F}(\delta x)$  denotes flow variables calculated with a mesh size of  $\delta x$.
From Fig. \ref{Fig-converge-Nu-Re}, we can see that the LB solver is second-order spatial accuracy for simulating velocity and temperature fields.
In addition, the vorticity field, which is obtained from the curl of velocity field data using a third-order finite difference scheme, also exhibits second-order spatial accuracy.
It should be noted that numerics in Table \ref{tab:converge} are presented with five significant numbers for clarity of writing;
while the data in Fig. \ref{Fig-converge-Nu-Re} are calculated with eight significant numbers.
The mesh independent tests indicate that a grid resolution of 256 is adequate for the simulation.
Thus, we use at least 256 grids to resolve the cell length or the cell height.
Other detailed simulation parameters are listed in Table \ref{tab:settings}.
In addition, we list the response parameters of $Nu$ and $Re$ in Table \ref{tab:settings}.
At fixed $Ra$ and $Pr$, with the increase in $\Gamma$, both the $Nu$ and $Re$ gradually increase to an asymptotic value.

\begin{table}[htbp]
\centering
\caption{Parameters of the different simulations (columns from left to right): Rayleigh number $Ra$, Prandlt number $Pr$, aspect ratio $\Gamma$, grid number, flow state (either steady state or unsteady state), average time $t_{avg}$ of the simulation in free-fall time unit $t_{f}=\sqrt{H/(g\beta_{T} \Delta_{T})}$ for unsteady flows, Nusselt number $Nu$, and Reynolds number $Re$.}
\begin{tabular}{ccccccccc}
\hline
$Ra$   &  $Pr$  & $\Gamma$  & Grid  & Flow state & $t_{avg}$ & $Nu$  & $Re$ \\
\hline
\ \ $1\times 10^{5}$ \ \ & \ \ 0.71 \ \ &  \ \ 0.75 \ \ & \ \ $256\times 342$ \ \ & \ \ Unsteady  \ \ & \ \ 1000 $t_{f}$  \ \ & \ \ 2.44 \ \ & \ \ 49.47 \ \ \\
\ \ $1\times 10^{5}$ \ \ & \ \ 0.71 \ \ &  \ \ 1 \ \    & \ \ $256\times 256$ \ \ & \ \ Steady    \ \ & \ \ -   \ \ & \ \ 3.92 \ \ & \ \ 84.43 \ \ \\
\ \ $1\times 10^{5}$ \ \ & \ \ 0.71 \ \ &  \ \ 2 \ \    & \ \ $512\times 256$ \ \ & \ \ Steady    \ \ & \ \ -   \ \ & \ \ 4.43 \ \ & \ \ 100.97 \ \ \\
\ \ $1\times 10^{5}$ \ \ & \ \ 0.71 \ \ &  \ \ 4 \ \    & \ \ $1024\times 256$ \ \ & \ \ Steady   \ \ & \ \ -   \ \ & \ \ 4.75 \ \ & \ \ 109.29 \ \ \\
\ \ $1\times 10^{5}$ \ \ & \ \ 0.71 \ \ &  \ \ 6 \ \    & \ \ $1536\times 256$ \ \ & \ \ Steady   \ \ & \ \ -  \ \  & \ \ 4.84 \ \ & \ \ 111.58 \ \ \\
\ \ $3\times 10^{5}$ \ \ & \ \ 0.71 \ \ &  \ \ 0.75 \ \ & \ \ $256\times 342$ \ \ & \ \ Steady    \ \ & \ \ -  \ \ & \ \ 3.34 \ \ & \ \ 99.93  \ \ \\
\ \ $3\times 10^{5}$ \ \ & \ \ 0.71 \ \ &  \ \ 1 \ \    & \ \ $256\times 256$ \ \ & \ \ Steady    \ \ & \ \ -  \ \ & \ \ 5.03 \ \ & \ \ 151.57 \ \ \\
\ \ $3\times 10^{5}$ \ \ & \ \ 0.71 \ \ &  \ \ 2 \ \    & \ \ $512\times 256$ \ \ & \ \ Steady    \ \ & \ \ -  \ \ & \ \ 5.73 \ \ & \ \ 182.68 \ \ \\
\ \ $3\times 10^{5}$ \ \ & \ \ 0.71 \ \ &  \ \ 4 \ \    & \ \ $1024\times 256$ \ \ & \ \ Steady   \ \ & \ \ -  \ \ & \ \ 6.22 \ \ & \ \ 199.36 \ \ \\
\ \ $3\times 10^{5}$ \ \ & \ \ 0.71 \ \ &  \ \ 6 \ \    & \ \ $1536\times 256$ \ \ & \ \ Steady   \ \ & \ \ -  \ \ & \ \ 6.37 \ \ & \ \ 203.89 \ \ \\
\ \ $1\times 10^{6}$ \ \ & \ \ 0.71 \ \ &  \ \ 0.75 \ \   & \ \ $256\times 342$ \ \ & \ \ Steady  \ \ & \ \ -  \ \  & \ \ 4.62 \ \ & \ \ 202.33 \ \ \\
\ \ $1\times 10^{6}$ \ \ & \ \ 0.71 \ \ &  \ \ 1 \ \   & \ \ $256\times 256$ \ \ & \ \ Steady   \ \ & \ \ -  \ \  & \ \ 6.33 \ \ & \ \ 275.70 \ \ \\
\ \ $1\times 10^{6}$ \ \ & \ \ 0.71 \ \ &  \ \ 2 \ \   & \ \ $512\times 256$ \ \ & \ \ Steady   \ \ & \ \ -  \ \  & \ \ 7.39 \ \ & \ \ 340.87 \ \ \\
\ \ $1\times 10^{6}$ \ \ & \ \ 0.71 \ \ &  \ \ 4 \ \   & \ \ $1024\times 256$ \ \ & \ \ Steady   \ \ & \ \ -  \ \  & \ \ 8.25 \ \ & \ \ 377.48 \ \ \\
\ \ $1\times 10^{6}$ \ \ & \ \ 0.71 \ \ &  \ \ 6 \ \   & \ \ $1536\times 256$ \ \ & \ \ Unsteady   \ \ & \ \ 1000 $t_{f}$  \ \  & \ \ 8.47 \ \ & \ \ 386.54 \ \ \\
\hline
\end{tabular}
\label{tab:settings}
\end{table}

\section{\label{Section3}Results and discussion}

\subsection{\label{Section3-1}General flow and heat transfer features}

\begin{figure}[htbp]
\centering
\includegraphics[width=16cm]{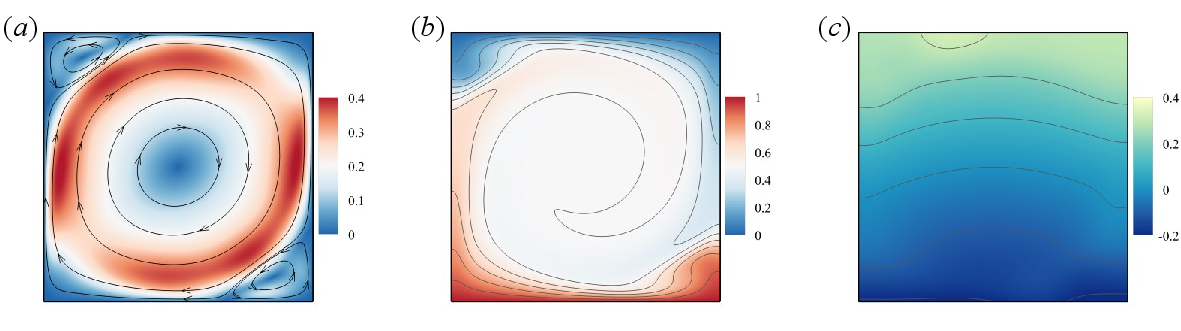}
\caption{\label{Fig-Ra1e6-Gamma1-flow} (\textit{a}) Contour of velocity magnitude (superposed with streamlines) $\sqrt{u^{*2}+v^{*2}}$, (\textit{b}) contour of temperature $T^{*}$, and (\textit{c}) contour of pressure $P^{*}$ at  $Ra=10^{6}$,  $Pr=0.71$, and  $\Gamma=1$.}
\end{figure}

We first present general flow and heat transfer patterns in the convection cell.
Figure \ref{Fig-Ra1e6-Gamma1-flow} shows the contour of velocity magnitude, temperature field, and pressure field in the  RB convection at  $Ra=10^{6}$,  $Pr=0.71$, and  $\Gamma=1$.
Here, the reference temperature is chosen as the temperature of the cold wall, that is,  $T_{0}=T_{cold}$ such that the dimensionless temperature varies between zero and one.
The reference pressure is chosen as the pressure at the cell center, that is, $P_{0}=P(x^{*}=0.5, y^{*}=0.5)$ such that the dimensionless pressure at the cell center equals zero.
In the RB convection, velocity magnitude exhibits large values near all four walls.
In the bulk region of the RB convection, the clockwise rotated large-scale vortex exhibits a circular shape; near the top-left and bottom-right corners, there exist two counterclockwise rotated small vortices [see Fig. \ref{Fig-Ra1e6-Gamma1-flow}(a)].
Thin thermal boundary layers appear near the bottom and top horizontal walls [see Fig. \ref{Fig-Ra1e6-Gamma1-flow}(b)], where the fluids are heating and cooling, respectively.
Driven by the clockwise large-scale circulation, detached hot plumes arise along the left adiabatic wall, and cold plumes fall along the right adiabatic wall.
From the pressure field [see Fig. \ref{Fig-Ra1e6-Gamma1-flow}(c)], we observe a strong gradient of pressure along the vertical direction, while the pressure contour is slightly twisted along the horizontal direction.
It should be noted that the pressure $P$ shown here (and the corresponding dimensionless one) is based on the Boussinesq approximation, and it is related to the hydrodynamic pressure $p_{N-S}$ via the relation of $P=p_{N-S}+\rho_{0}gy$.
For the dimensionless pressure under the Boussinesq approximation $P^{*}=(P-P_{0})/(\rho_{0}g\beta_{T}\Delta_{T}H)$, we can rewrite it as $P^{*}=[p_{N-S}-(p_{N-S})_{0}]/(\rho_{0}g\beta_{T}\Delta_{T}H)+(y^{*}-0.5)/(\beta_{T}\Delta_{T})$, which indicates there is an additional term $(y^{*}-0.5)/(\beta_{T}\Delta_{T})$ compared to that in the dimensionless hydrodynamic pressure $p_{N-S}^{*}=[p_{N-S}-(p_{N-S})_{0}]/(\rho_{0}g\beta_{T}\Delta_{T}H)$.
%%%Thus, under the Boussinesq approximation, the dimensionless pressure $P^{*}$ is higher near the top wall due to positive elevation.

\begin{figure}[htbp]
\centering
\includegraphics[width=8cm]{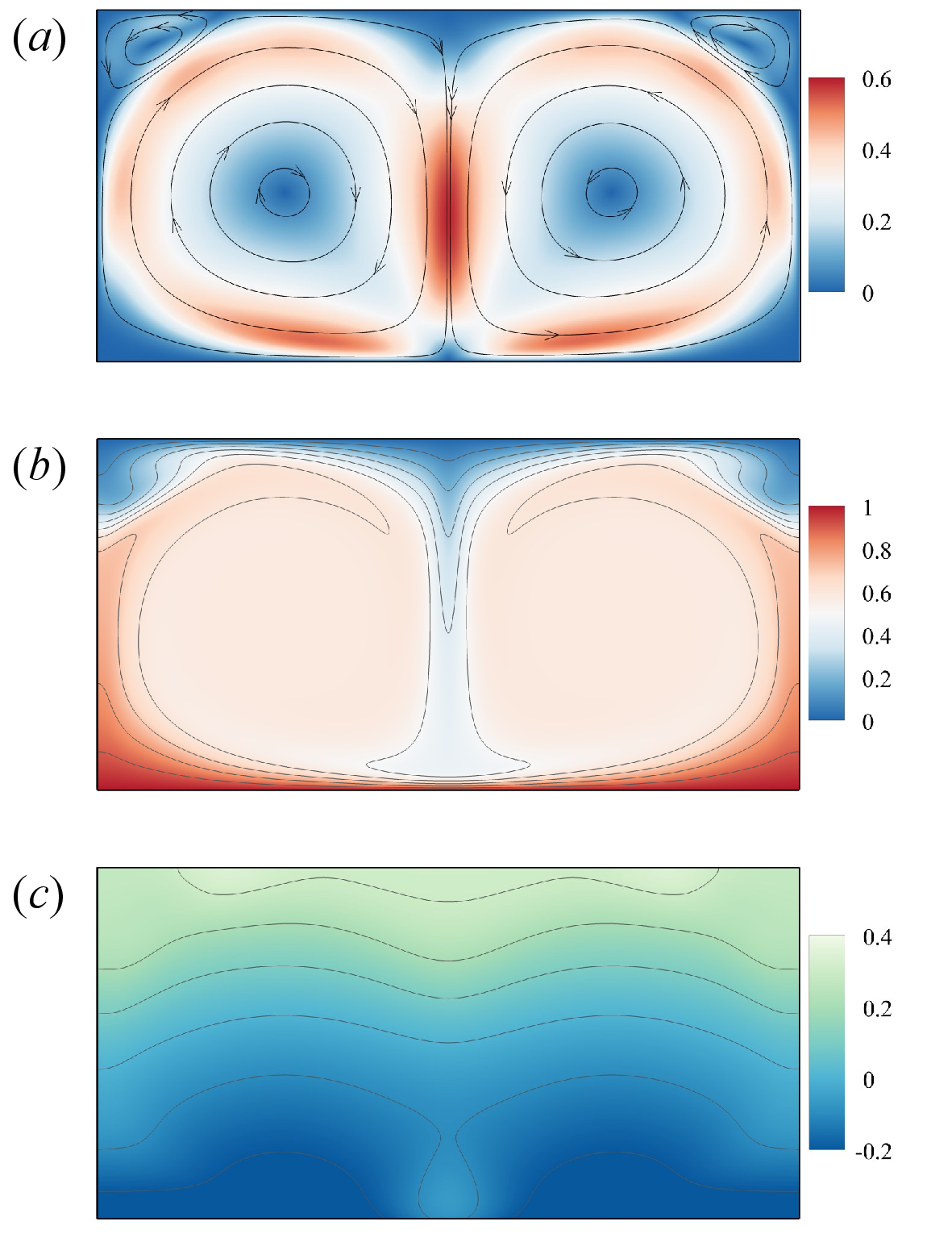}
\caption{\label{Fig-Ra1e6-Gamma2-flow} (\textit{a}) Contour of velocity magnitude (superposed with streamlines) $\sqrt{u^{*2}+v^{*2}}$, (\textit{b}) contour of temperature $T^{*}$, (\textit{c}) contour of pressure $P^{*}$ at $Ra=10^{6}$,  $Pr=0.71$, and $\Gamma=2$.}
\end{figure}

In the above, we discussed flow and heat transfer patterns in the convection cell of a unit square, that is, cell aspect ratio $\Gamma=1$.
Next, we focus on rectangular cells with  $\Gamma>1$ and  $\Gamma<1$, respectively.
In Fig. \ref{Fig-Ra1e6-Gamma2-flow}, we show flow and heat transfer patterns in the RB convection with  $\Gamma=2$.
We can observe two horizontally stacked main vortices in the $\Gamma=2$  cell [see Fig. \ref{Fig-Ra1e6-Gamma2-flow}(a)].
These two counter-rotating vortices carry hot rising plumes along the vertical left and right walls.
At the top wall, the fluids are cooling down, cold plumes emerge and they fall along the vertical mid-plane of the cell, and the temperature field preserves left-right symmetry, as shown in Fig. \ref{Fig-Ra1e6-Gamma2-flow}(b).
The contour of the pressure field in the RB convection [see Fig. \ref{Fig-Ra1e6-Gamma2-flow}(c)] is also twisted along the horizontal direction, and the trend is more obvious where the falling cold plumes eject on the hot bottom walls.

\begin{figure}[htbp]
\centering
\includegraphics[width=16cm]{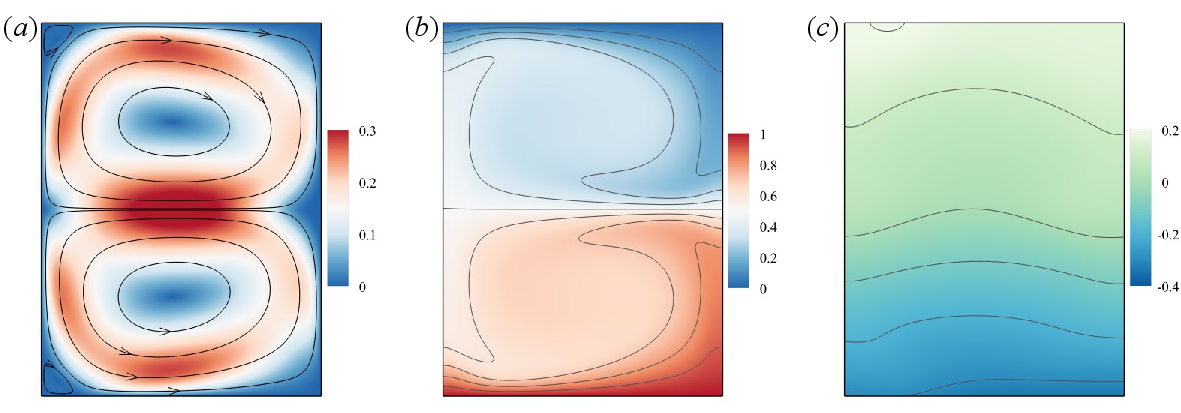}
\caption{\label{Fig-Ra1e6-Gamma0.75-flow} (\textit{a}) Contour of velocity magnitude (superposed with streamlines) $\sqrt{u^{*2}+v^{*2}}$, (\textit{b}) contour of temperature $T^{*}$, (\textit{c}) contour of pressure $P^{*}$ at  $Ra=10^{6}$,  $Pr=0.71$, and  $\Gamma=0.75$.}
\end{figure}

Figure \ref{Fig-Ra1e6-Gamma0.75-flow} further presents flow and heat transfer patterns in the RB convection with $\Gamma=0.75$.
We observe two vertically stacked main vortices in the  $\Gamma=0.75$ cell [see Fig. \ref{Fig-Ra1e6-Gamma0.75-flow}(a)].
The velocity magnitude reaches its maximum at the horizontal mid-plane, where the two main vortices interact.
The bottom counterclockwise rotated vortex carries hot rising plumes, while the top clockwise rotated vortex carries cold falling plumes [see Fig. \ref{Fig-Ra1e6-Gamma0.75-flow}(b)].
The plumes exchange thermal energy at the horizontal mid-plane, and the temperature field preserves the top-bottom symmetry.
In addition, the contour of the pressure field along the horizontal direction in the $\Gamma=0.75$  cell [see Fig. \ref{Fig-Ra1e6-Gamma0.75-flow}(c)] is less twisted than that in the $\Gamma=2$  cell [see Fig. \ref{Fig-Ra1e6-Gamma2-flow}(c)].

\subsection{\label{Section3-4}Vorticity production due to wall shear stress}

\begin{figure}[htbp]
\centering
\includegraphics[width=14cm]{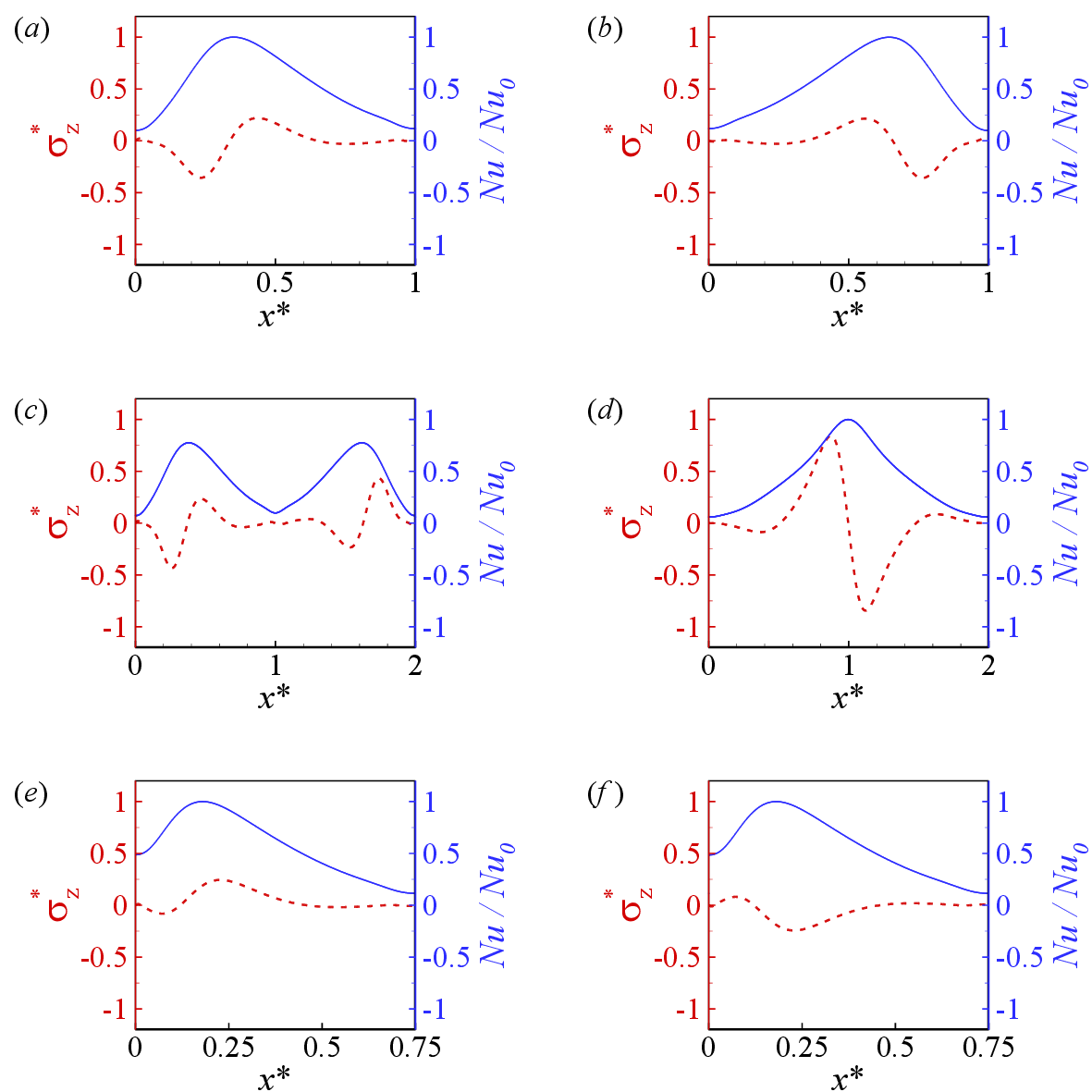}
\caption{\label{Fig-vorFlux} Distribution of dimensionless vorticity flux  $\sigma_{i}^{*}=-\sqrt{Pr/Ra}\cdot n_{j^{*}}\partial_{j^{*}}\omega_{i}^{*}$ (red dash-dot line) and normalized Nusselt number $Nu/Nu_{0}$ (blue solid line) along the top cooling wall (left-column) and bottom heating wall (right-column)  at  $Ra=10^{6}$,  $Pr=0.71$,  (\textit{a}) and (\textit{b}) $\Gamma=1$, (\textit{c}) and  (\textit{d}) $\Gamma=2$, (\textit{e}) and (\textit{f}) $\Gamma=0.75$.}
\end{figure}

To describe how much vorticity is entering the flow from the wall, we used the metrics of vorticity flux  $\sigma_{i}=-\nu n_{j}\partial_{j}\omega_{i}$.
Here,  $n_{i}$ is the plane orientation.
In two dimensions, we only have non-zero components of vorticity as  $\omega_{z}=\partial v/\partial x-\partial u/\partial y$.
Thus, the vorticity flux along the vertical walls (i.e., $n_{y}=0$) is $\sigma_{z}=-\nu n_{x}\partial_{x}\omega_{z}$, and the vorticity flux along the horizontal walls (i.e., $n_{x}=0$) is  $\sigma_{z}=-\nu n_{y}\partial_{y}\omega_{z}$.
On the other hand, the heat flux perpendicular to the wall is  $q_{i}=-\kappa\partial_{i}T$, and we normalize the heat flux with that due to pure conduction $\kappa\Delta_{T}H$, and then, we can use the dimensionless Nusselt number at the wall $Nu=q_{i}/(\kappa\Delta_{T}/H)=-\partial_{i^{*}}T^{*}$ to describe how much thermal energy is entering the flow.
Here, $\kappa$ denotes the thermal conductivity of the fluid.
In Fig. \ref{Fig-vorFlux}, we plot the distribution of dimensionless vorticity flux  $\sigma_{i}^{*}=-\sqrt{Pr/Ra}\cdot n_{j^{*}}\partial_{j^{*}}\omega_{i}^{*}$ and the normalized Nusselt number $Nu/Nu_{0}$  along top and bottom walls.
Here,  $Nu_{0}$ denotes the maximum value of $Nu$ along all four walls.
In the  $\Gamma=1$ cell [see Figs. \ref{Fig-vorFlux}(a) and \ref{Fig-vorFlux}(b)], the spatial distribution of the $\sigma_{i}^{*}$ profile and the $Nu$ profile is antisymmetric along the top and bottom walls.
Along the horizontal walls, we can observe one peak and one valley near the central part of the wall.
Here, the negative values of $\sigma_{i}^{*}$  imply the vorticity that enters the flow is clockwise rotating.
Both the positive and negative neighboring vorticity lead to enhanced thermal energy that enters the flow, and thus, we can observe a peak in the $Nu$.

In the $\Gamma=2$  cell [see Figs. \ref{Fig-vorFlux}(c) and \ref{Fig-vorFlux}(d)], along the top (or the bottom) wall, the minimum values of $Nu$ correspond to the positions where cold (or hot) plumes are released, and the maximum values of $Nu$ correspond to the positions where hot (or cold) plumes are imping and penetrating the wall.
Previously, Kenjere{\v{s}} and Hanjali{\'c} \cite{kenjerevs2000convective} reported the asymmetric of the $Nu$ profile along the top and bottom walls in the $\Gamma>1$ cell.
Here, we found that the $\sigma_{i}^{*}$ profile along the top and bottom walls is also asymmetric.
More importantly, we found the spatial distribution of $Nu$ along the heating and cooling walls can be well explained by the spatial distribution of $\sigma_{i}^{*}$.
In the $\sigma_{i}^{*}$ profile, a peak and a valley imply vorticity with opposite rotating direction is entering the flow; correspondingly,
we can then observe a peak in the $Nu$ profile.
The reason is that the amount of vorticity (regardless of the rotating direction) that enters the flow is directly related to the amount of thermal energy that enters the flow.
In Fig. \ref{Fig-vorFlux}(c), along the top wall, there are two pairs of peak and valley in the vorticity flux profile, and thus, there are two peaks in the $Nu$ profile.
In Fig. \ref{Fig-vorFlux}(d), along the bottom wall, there is one peak and one valley in the vorticity flux profile, and thus, there is only one peak in the $Nu$ profile.
In the $\Gamma=0.75$  cell [see Figs. \ref{Fig-vorFlux}(e) and \ref{Fig-vorFlux}(f)], the two large-scale vortices are vertically stacked, and there is only one peak and valley in the  $\sigma_{i}^{*}$ profile along the horizontal walls, which is similar to that in the $\Gamma=1$  cell.
Correspondingly, the $Nu$ profile shows a peak where extreme values of $\sigma_{i}^{*}$  appear.
Thus, we can analyze the spatial distribution of $Nu$ along the heating and cooling walls via the vorticity flux along the wall.

\begin{figure}[htbp]
\centering
\includegraphics[width=16cm]{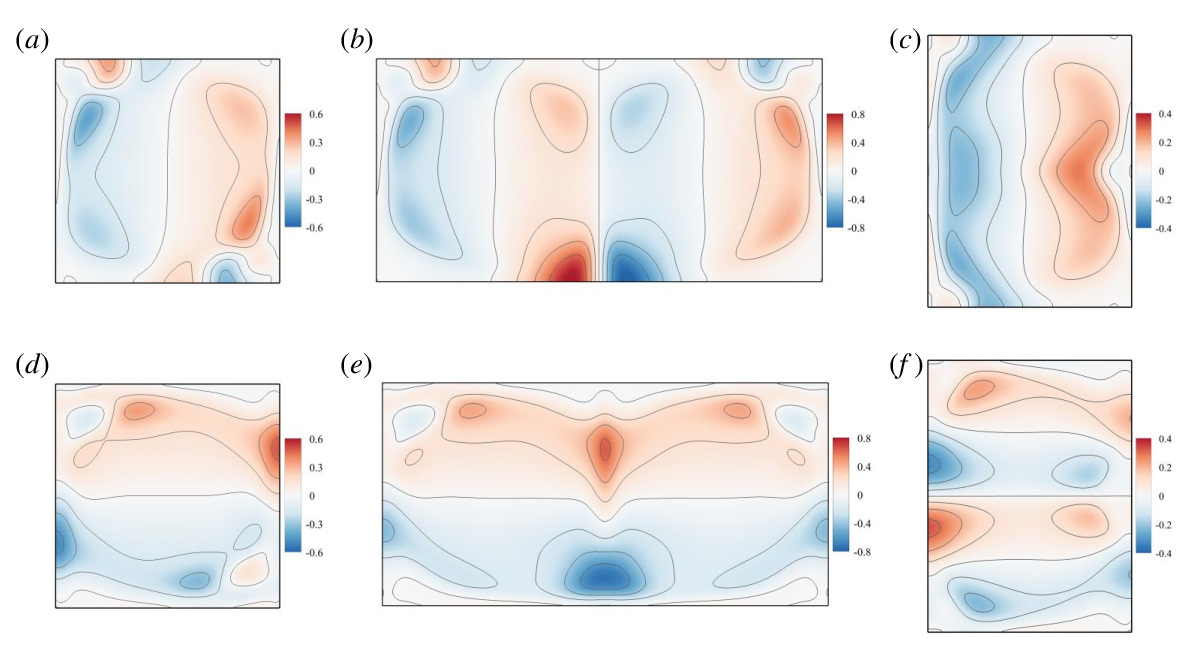}
\caption{\label{Fig-pressureGrad-RBC} The [top panel, (\textit{a})-(\textit{c})] $x$-direction component, [bottom panel, (\textit{d})-(\textit{f})] $y$-direction component of pressure gradients  at  $Ra=10^{6}$,  $Pr=0.71$,  (\textit{a}) and (\textit{d}) $\Gamma=1$, (\textit{b}) and (\textit{e}) $\Gamma=2$, (\textit{c}) and (\textit{f}) $\Gamma=0.75$.}
\end{figure}

Although vorticity can be viewed as the signatures of fluid motion, it is the pressure gradients that drive the flow.
In the vorticity transport equation,  the pressure does not appear; thus, it does not directly change the vorticity.
It should be noted that the buoyancy appears in the momentum equation as an external body force, and thus, the effective pressure gradients are a combination of original pressure gradients and buoyancy.
In Fig. \ref{Fig-pressureGrad-RBC}, we show the dimensionless effective pressure gradients (i.e., $\nabla P_{x}^{*}$  and  $\nabla P_{y}^{*}-T^{*}$) in the RB convection.
We are particularly interested in the pressure gradients along the wall, which is necessary to sustain a flux of vorticity into the fluid.
Because the velocity is zero at the wall, from the momentum equation, we have  $\nabla P^{*}-T^{*}\hat{\mathbf{y}}=-\sqrt{Pr/Ra}\nabla \times\mathbf{\omega}^{*}$.
Comparing Fig. \ref{Fig-vorFlux} with Fig. \ref{Fig-pressureGrad-RBC}, we can verify that positive $\sigma_{i}^{*}$  along the bottom (or top) wall is indeed caused by the positive (or negative) value of horizontal effective pressure gradient.

\begin{figure}[htbp]
\centering
\includegraphics[width=15cm]{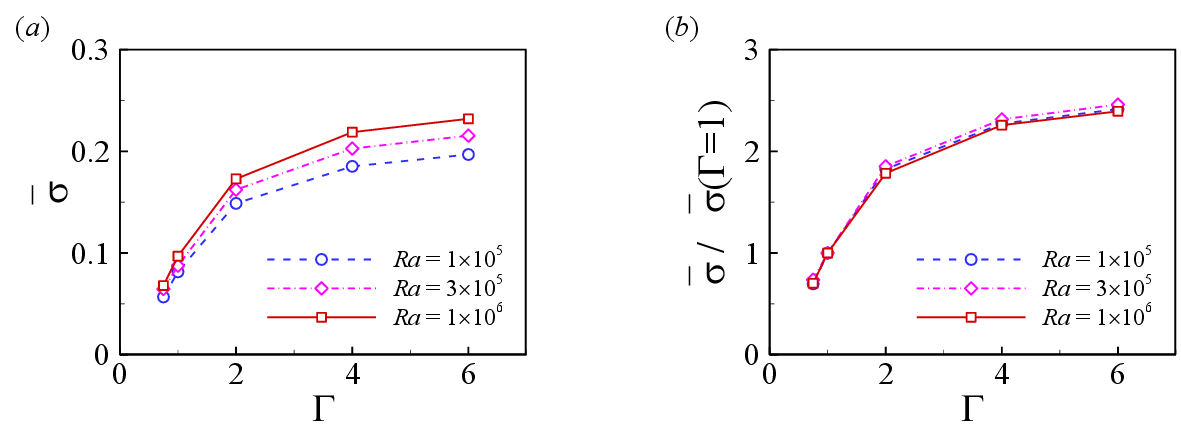}
\caption{\label{Fig-vorFluxGamma} Average dimensionless vorticity flux  $\sigma_{i}^{*}=-\sqrt{Pr/Ra}\cdot n_{j^{*}}\partial_{j^{*}}\omega_{i}^{*}$ along the top and bottom walls: (\textit{a}) $\bar{\sigma}=(\langle |\sigma_{z}^{*}| \rangle_{top}+\langle |\sigma_{z}^{*}| \rangle_{bottom})/2$, (\textit{b}) normalized $\bar{\sigma}/\bar{\sigma}(\Gamma=1)$ at  $10^{5} \le Ra \le 10^{6}$,  $Pr=0.71$, and $0.75 \le \Gamma \le 6$.}
\end{figure}

To quantitatively describe the vorticity flux as functions of the flow control parameters, we calculate the average of $\sigma_{i}^{*}$ along the horizontal top and bottom walls.
Because positive and negative values of $\sigma_{i}^{*}$  simply imply the vorticity that enters the flow with opposite rotating direction, we take the absolute values of $\sigma_{i}^{*}$ to do the average, namely, $\bar{\sigma}=(\langle |\sigma_{z}^{*}| \rangle_{top}+\langle |\sigma_{z}^{*}| \rangle_{bottom})/2$.
Figure \ref{Fig-vorFluxGamma} (a) shows the aspect ratio dependence of average vorticity flux  $\bar{\sigma} $ at  $10^{5} \le Ra \le 10^{6}$,  $Pr=0.71$, and $0.75 \le \Gamma \le 6$.
We can see that with the increase in $\Gamma$, the average vorticity flux along the horizontal walls and monotonously increases and gradually approach an asymptotic value.
We then normalize $\bar{\sigma}$ with $\bar{\sigma}(\Gamma=1)$, and all data fall on top of each other and the shape of $\bar{\sigma}$ do not change with $Ra$, as shown in Fig. \ref{Fig-vorFluxGamma} (b).
The results shown here thus suggest universal properties of vorticity flux along the heating and cooling walls with respect to different $Ra$.

\subsection{Vorticity production due to buoyancy}

\begin{figure}[htbp]
\centering
\includegraphics[width=16cm]{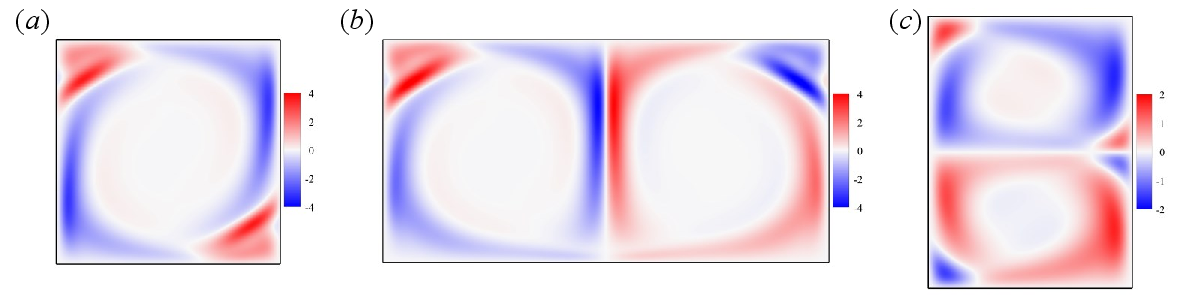}
\caption{\label{Fig-vorBodyForce} Contour of vorticity produced by buoyancy, i.e., $(\nabla \times \mathbf{F})^{*}$, at  $Ra=10^{6}$,  $Pr=0.71$ and  (\textit{a}) $\Gamma=1$, (\textit{b}) $\Gamma=2$, and (\textit{c}) $\Gamma=0.75$.}
\end{figure}

The vorticity transport equation for incompressible thermal convection can be obtained by taking the curl of the momentum equation, which is written as \cite{wu2007vorticity}
\begin{equation}
\frac{\partial \mathbf{\omega}}{\partial t}+
\underbrace{\mathbf{u}\cdot\nabla\mathbf{\omega}}_{\text{vorticity advection}}
=\underbrace{\mathbf{\omega}\cdot \nabla\mathbf{u}}_{\text{vortex stretching/tilting}}
+\underbrace{\nu\nabla^{2}\mathbf{\omega}}_{\text{vorticity diffusion}}
+\underbrace{\nabla\times \mathbf{F}}_{\text{body force produced vorticity}}
\label{Eq.vorticity}
\end{equation}
In two dimensions, the velocity vector is $\mathbf{u}=(u,v,0)$, and thus, we only have non-zero components of vorticity as $\omega_{z}=\partial v/\partial x-\partial u/\partial y$.
There is no vortex stretching nor tilting ($\omega \cdot \nabla \mathbf{u}$ is always zero), and thus, vorticity production does not come from shearing effects in two dimensions.
In addition, the body force (i.e., buoyancy) is $\mathbf{F}=[0,g\beta_{T}(T-T_{0}),0]$; then, the above vorticity transport equation  can be rewritten as
\begin{equation}
\frac{\partial \mathbf{\omega}_{z}}{\partial t}+\mathbf{u} \cdot \nabla \mathbf{\omega}_{z}
=\nu \nabla^{2}\mathbf{\omega}_{z}
+g\beta_{T}\frac{\partial T}{\partial x}
\end{equation}
With the non-dimensional group in Eq. (\ref{Eq.non-dimensional}), the dimensionless vorticity transport equation for incompressible thermal flows is written as
\begin{equation}
\frac{\partial \omega_{z}^{*}}{\partial t^{*}}+
\mathbf{u}^{*} \cdot \nabla \mathbf{\omega}_{z}
=\sqrt{\frac{Pr}{Ra}}\nabla^{2}\mathbf{\omega}_{z}^{*}
+\frac{\partial T^{*}}{\partial x^{*}}
\label{Eq.2Dvorticity}
\end{equation}
In Fig. \ref{Fig-vorBodyForce}, we further provide the contour of vorticity produced by buoyancy, namely, the contour of $(\nabla \times \mathbf{F})^{*}$.
An interesting finding is that the position of buoyancy-produced vorticity accumulates exactly where the vortices lay.
Specifically, in the $\Gamma=1$  cell [see Fig. \ref{Fig-vorBodyForce}(a)], counterclockwise rotated buoyancy-produced vorticity accumulates near the left-top and right-bottom corner of the cell, where counterclockwise rotated corner vortices exist [compared to Fig. \ref{Fig-Ra1e6-Gamma1-flow}(a)].
In addition, clockwise rotated buoyancy-produced vorticity accumulates and forms the edge of the main clockwise rotating vortex.
Thus, in Fig. \ref{Fig-vorBodyForce}, the red color region represents counterclockwise rotating vortices, and the blue color region represents the clockwise rotating vortices.
We can also observe similar patterns in the $\Gamma>1$  and the $\Gamma<1$ cell.
We propose to use the contour of buoyancy-produced vorticity, that is, $(\nabla \times \mathbf{F})^{*}$, as a metric to visualize the vortex in the RB convection.
It should be mentioned that at present, this approach is only valid for two-dimensional flow, and extension to three-dimensional flows deserves future investigation.

\begin{figure}[htbp]
\centering
\includegraphics[width=16cm]{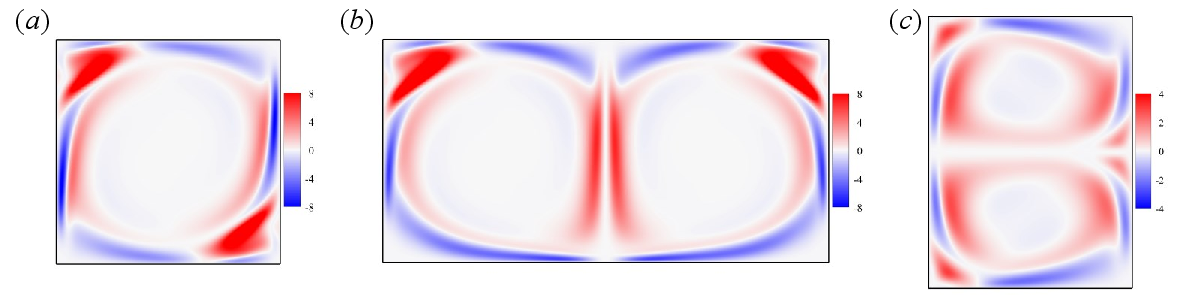}
\caption{\label{Fig-enstrophy} Contour of enstrophy produced by buoyancy, that is, $\left[\mathbf{\omega}\cdot (\nabla \times \mathbf{F})\right]^{*}$, at  $Ra=10^{6}$,  $Pr=0.71$ and  (\textit{a}) $\Gamma=1$, (\textit{b}) $\Gamma=2$, and (\textit{c}) $\Gamma=0.75$.}
\end{figure}

We also check the amount of vorticity, without regard to direction, which is measured by enstrophy $\omega^{2}/2$.
The enstrophy transport equation can be obtained via the inner production of $\mathbf{\omega}$ with Eq. (\ref{Eq.vorticity}),
which is written as \cite{wu2007vorticity}
\begin{equation}
\frac{\partial \left(\frac{1}{2}\omega^{2} \right)}{\partial t}
+\mathbf{u}\cdot \nabla\left(\frac{1}{2}\omega^{2} \right)
=\mathbf{\omega}\cdot \mathbf{\omega}\cdot \mathbf{S}
+\nu\nabla^{2}\left(\frac{1}{2}\omega^{2} \right)
-\nu\left( \nabla \omega\right)^{2}
+\mathbf{\omega}\cdot \left(\nabla \times \mathbf{F} \right)
\end{equation}
Here, $\mathbf{S}=(\partial_{j}u_{i}+\partial_{i}u_{j})/2$ denotes the strain rate tensor.
In two dimensions, the term $\mathbf{\omega}\cdot \mathbf{\omega}\cdot \mathbf{S}$ is always zero.
With the non-dimensional group in Eq. (\ref{Eq.non-dimensional}), the dimensionless enstrophy transport equation for incompressible thermal flows in two dimensions is written as
\begin{equation}
\frac{\partial \left(\frac{1}{2}\omega_{z}^{*2} \right)}{\partial t}
+\mathbf{u}^{*}\cdot \nabla \left(\frac{1}{2}\omega_{z}^{*2} \right)
=\sqrt{\frac{Pr}{Ra}} \nabla^{2}\left( \frac{1}{2}\omega_{z}^{*2} \right)
-\sqrt{\frac{Pr}{Ra}} \left( \nabla \omega_{z}^{*} \right)^{2}
+\omega_{z}^{*} \cdot \frac{\partial T^{*}}{\partial x^{*}}
\end{equation}
In Fig. \ref{Fig-enstrophy}, we provide the contour of enstrophy produced by buoyancy.
We can see that due to buoyancy, intense enstrophy production (or destruction) occurs at the inner (or outer) edge of the main vortices; intense enstrophy production also occurs at regions of corner vortices.
Analysis of spatial distribution of buoyancy-produced vorticity and buoyancy-produced enstrophy indicate that buoyancy significantly affects the flow structure in the Rayleigh-B\'enard convection,
and contour of buoyancy-produced vorticity and buoyancy-produced enstrophy can be used to visualize the vortices in the two-dimensional Rayleigh-B\'enard convection.

\begin{figure}[htbp]
\centering
\includegraphics[width=15cm]{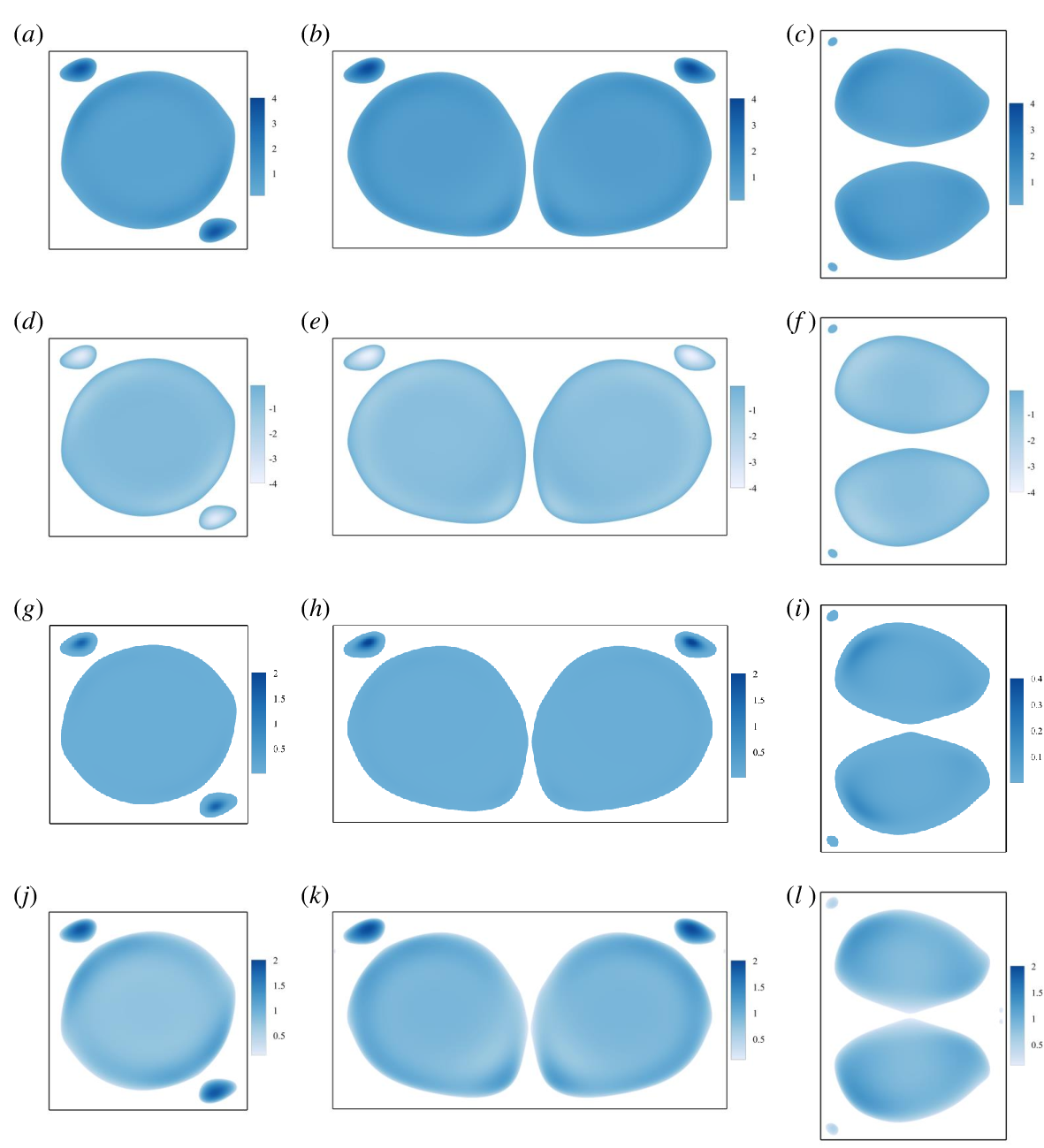}
\caption{\label{Fig-vorIdentification} Vortex identification using (\textit{a})-(\textit{c}) the Q-criterion (contour of $Q>0.1$), (\textit{d})-(\textit{f}) the $\lambda_{2}$-criterion (contour of $\lambda_{2}<-0.1$), (\textit{g})-(\textit{i}) the  $\Delta$-criterion (contour of $\Delta>10^{-6}$), (\textit{j})-(\textit{l}) the swirling-strength criterion (contour of $\lambda_{c_{i}}>0.1$) at $Ra=10^{6}$, $Pr=0.71$,  and  $\Gamma=1$ (left-column),  $\Gamma=2$ (middle-column),  $\Gamma=0.75$ (right-column).}
\end{figure}

It should be noted that the precise definition of a vortex is still an open question.
Previously, various criteria to identify a vortex have been proposed. \cite{jcr1988eddies,jeong1995identification,chong1990general,zhou1999mechanisms}
In Fig. \ref{Fig-vorIdentification}, we show the vortex identification using
the Q-criterion, \cite{jcr1988eddies}
the $\lambda_{2}$-criterion, \cite{jeong1995identification}
the $\Delta$-criterion, \cite{chong1990general}
and the swirling-strength criterion \cite{zhou1999mechanisms}
in the RB convection.
Although those methods can be used to identify the vortices region, they failed to indicate the rotating direction of the vortices.
Our method can be used to identify the vortices regions as well as the rotating direction of the vortices, which serves as an efficient vortex visualization approach.
Overall, our analysis of the vorticity distribution provides an alternative understanding of the flow structure in the thermal convection cells.

\subsection{Vorticity transport due to advection and diffusion}

\begin{figure}[htbp]
\centering
\includegraphics[width=16cm]{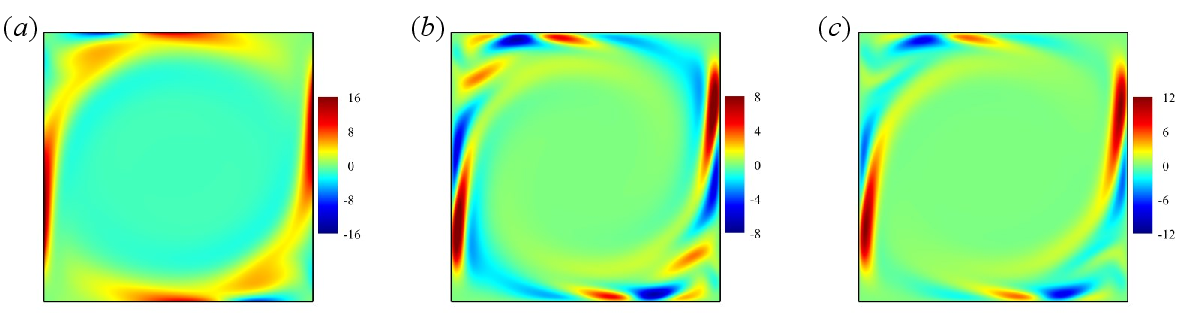}
\caption{\label{Fig-Ra1e6-Gamma1-vor} Contour of (\textit{a}) vorticity  $\omega_{z}^{*}$, (\textit{b}) vorticity advection $\mathbf{u}^{*}\cdot\nabla \omega_{z}^{*}$, (\textit{c}) vorticity diffusion $\sqrt{Pr/Ra}\nabla^{2}\omega_{z}^{*}$  at  $Ra=10^{6}$,  $Pr=0.71$, and  $\Gamma=1$.}
\end{figure}

The vorticity redistribution mechanism is mainly due to advection and diffusion transport.
In Fig. \ref{Fig-Ra1e6-Gamma1-vor}(a), we show the contour of the vorticity component $\omega_{z}^{*}$ in the $\Gamma=1$ cell.
We can observe strong positive values of $\omega_{z}^{*}$ (i.e., counterclockwise rotated vorticity) mainly appear near the walls, and it is near zero in the bulk region of the cell.
This can be simply understood as the velocity gradients are large near the wall, as evident from the clockwise rotated large-scale circulation "wind" in the cell [see Fig. \ref{Fig-Ra1e6-Gamma1-flow}(a)].
A fluid particle near the left bottom (or the right top) wall will first feel the influence of the large-scale circulation and gain some vorticity.
After some time, it will move up (or down) along the vertical walls and diffuse out from the wall; thus, the thickness of the diffusion layer for vorticity near the left (or the right) wall will grow from the bottom (or top) to top (or bottom).
In Fig. \ref{Fig-Ra1e6-Gamma1-vor}(b), we further show advection of vorticity $\mathbf{u}^{*}\cdot\nabla \omega_{z}^{*}$ and we can see that two pairs of counter-rotating vorticity coexist along each of the four walls.
Near the wall, the advection of vorticity [see Fig. \ref{Fig-Ra1e6-Gamma1-vor}(b)] is strongly influenced by the redistribution of wall-produced vorticity.
Taking the top wall as an example, at around $0.15 < x^{*} < 0.3$, negative values of vorticity are entering the fluid [as evident by the valley in vorticity flux profile shown in Fig. \ref{Fig-vorFlux}(a), see more discussion in Sec. \ref{Section3-4}], which corresponds to the advection of clockwise rotated vorticity [i.e., a negative value in the contour of Fig. \ref{Fig-Ra1e6-Gamma1-vor}(b)].
On the contrary, at around $0.35 < x^{*} < 0.5$, positive values of vorticity are entering the fluid [as evident by the peak in vorticity flux profile shown in Fig. \ref{Fig-vorFlux}(a)], which corresponds to the advection of counterclockwise rotated vorticity [i.e., positive value in the contour of Fig. \ref{Fig-Ra1e6-Gamma1-vor}(b)].
Similarly, we can analyze the distribution of vorticity advection near the other three walls.
As for the diffusion of vorticity due to the viscous effect, the pattern of vorticity diffusion is similar to that of vorticity advection [see Fig. \ref{Fig-Ra1e6-Gamma1-vor}(c)], and the main differences are caused by the production of either counterclockwise or clockwise rotated vorticity due to buoyancy.
Meanwhile, from Fig. \ref{Fig-Ra1e6-Gamma1-vor}, we can see that in the $\Gamma=1$ cell, the redistribution of vorticity advection and  vorticity diffusion preserves top-bottom antisymmetric property.
In the RB convection with $\Gamma=2$ or $\Gamma=0.75$, the aspect ratio does not significantly influence the vorticity transport.
For the sake of clarity, we will not repeat the detailed discussion here, but only show the contour plot in the Appendix.

\section{\label{Section4}Conclusions}
In this work, we analyzed the vorticity production and transport in the RB convection.
Specifically, with flow field and temperature field information (via solving fluid flows and heat transfer equations), we calculated each term in the vorticity transport equation.
We examined the vorticity advection, diffusion, and production due to buoyancy and wall shear stress.
The main findings are summarized as follows:
\begin{enumerate}

\item
The flow structure and temperature distribution vary with  $\Gamma$ greatly due to multiple vortices interaction.
In addition to vorticity redistribution, buoyancy produces significant vorticity in the bulk region, resulting in a richer vortices structure.

\item
The main vortices and the corner vortices can be visualized via the contour of buoyancy-produced vorticity.
This method can be used to identify the vortices regions as well as the rotating direction of the vortices, which serves as an efficient vortex visualization approach.

\item
The spatial distribution of vorticity flux along the heating and cooling walls is positively correlated with that of $Nu$, suggesting the amount of vorticity that enters the flow is directly related to the amount of thermal energy that enters the flow.
\end{enumerate}

\begin{acknowledgments}
This work was supported by the National Natural Science Foundation of China (NSFC) through Grant Nos. 11902268 and 11772259,
the National Key Project No. GJXM92579,
the Fundamental Research Funds for the Central Universities of China (No. D5000200570),
and the 111 project of China (No. B17037).
\end{acknowledgments}

\section*{AUTHOR DECLARATIONS}
\subsection*{Conflict of Interest}
The authors declare no competing interests.

\section*{Data Availability}
The data that support the findings of this study are available from the corresponding author upon reasonable request.

\section*{Appendix: Vorticity transport in $\Gamma=2$ cell and $\Gamma=0.75$ cell}
\begin{figure}[htbp]
\centering
\includegraphics[width=8cm]{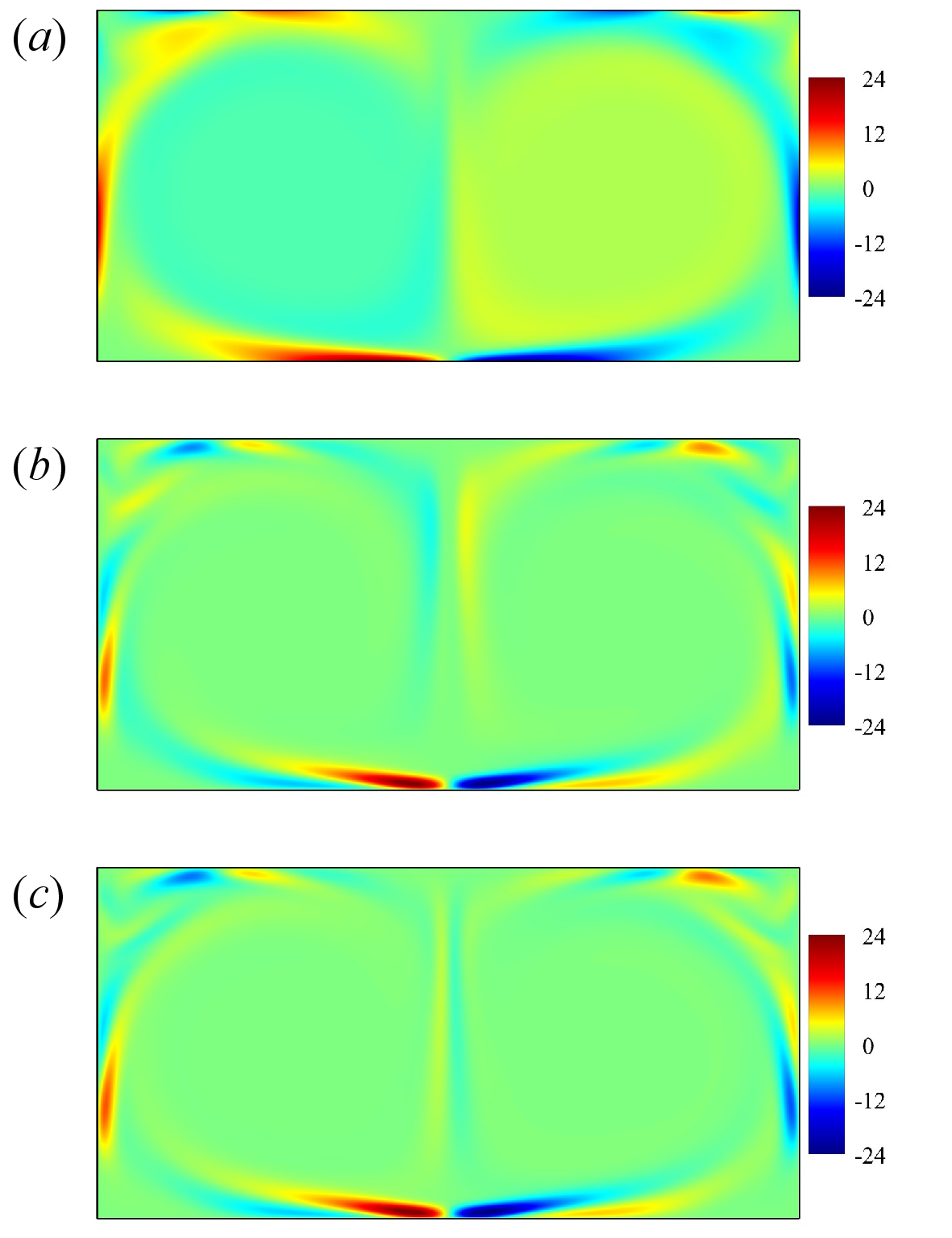}
\caption{\label{Fig-Ra1e6-Gamma2-vor} Contour of (\textit{a}) vorticity  $\omega_{z}^{*}$, (\textit{b})  vorticity advection $\mathbf{u}^{*}\cdot\nabla \omega_{z}^{*}$, (\textit{c}) vorticity diffusion $\sqrt{Pr/Ra}\nabla^{2}\omega_{z}^{*}$  at  $Ra=10^{6}$,  $Pr=0.71$, and  $\Gamma=2$.}
\end{figure}

\begin{figure}[htbp]
\centering
\includegraphics[width=16cm]{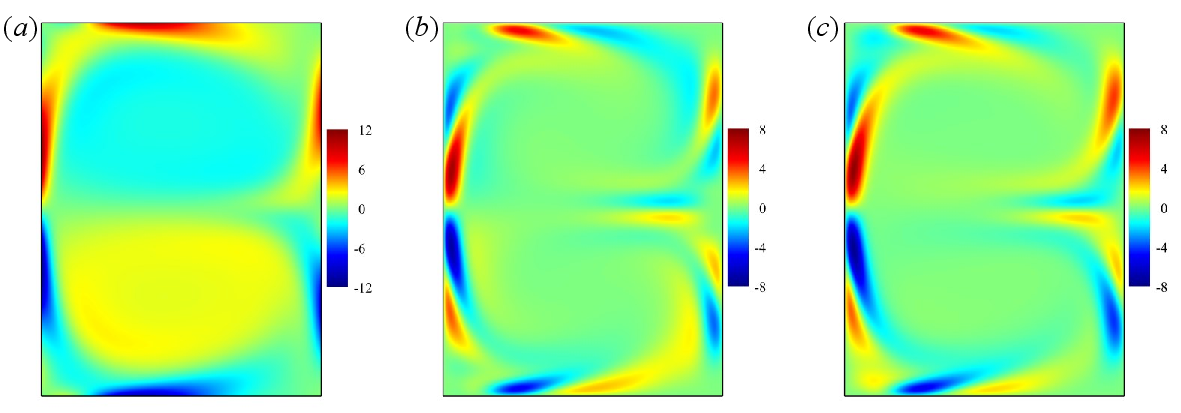}
\caption{\label{Fig-Ra1e6-Gamma0.75-vor} Contour of (\textit{a}) vorticity  $\omega_{z}^{*}$, (\textit{b}) vorticity advection $\mathbf{u}^{*}\cdot\nabla \omega_{z}^{*}$, (\textit{c}) vorticity diffusion $\sqrt{Pr/Ra}\nabla^{2}\omega_{z}^{*}$  at  $Ra=10^{6}$,  $Pr=0.71$, and  $\Gamma=0.75$.}
\end{figure}

In the RB convection with $\Gamma=2$, the clockwise (or counterclockwise) rotated main vortex sitting in the left-half (or right-half) of the cell induce counterclockwise (or clockwise) rotated vorticity near the horizontal walls (see Fig. \ref{Fig-Ra1e6-Gamma2-vor}).
In the RB convection with  $\Gamma=0.75$, the clockwise (or counterclockwise) rotated main vortex sitting in the top-half (or bottom-half) of the cell induce counterclockwise (or clockwise) rotated vorticity near the vertical walls (see Fig. \ref{Fig-Ra1e6-Gamma0.75-vor}).
As for the differences among vorticity, vorticity advection, and vorticity diffusion, one may refer to the previous discussion in the $\Gamma=1$ cell.


\nocite{*}
\bibliography{myBib}% Produces the bibliography via BibTeX.

\begin{thebibliography}{37}%
\makeatletter
\providecommand \@ifxundefined [1]{%
 \@ifx{#1\undefined}
}%
\providecommand \@ifnum [1]{%
 \ifnum #1\expandafter \@firstoftwo
 \else \expandafter \@secondoftwo
 \fi
}%
\providecommand \@ifx [1]{%
 \ifx #1\expandafter \@firstoftwo
 \else \expandafter \@secondoftwo
 \fi
}%
\providecommand \natexlab [1]{#1}%
\providecommand \enquote  [1]{``#1''}%
\providecommand \bibnamefont  [1]{#1}%
\providecommand \bibfnamefont [1]{#1}%
\providecommand \citenamefont [1]{#1}%
\providecommand \href@noop [0]{\@secondoftwo}%
\providecommand \href [0]{\begingroup \@sanitize@url \@href}%
\providecommand \@href[1]{\@@startlink{#1}\@@href}%
\providecommand \@@href[1]{\endgroup#1\@@endlink}%
\providecommand \@sanitize@url [0]{\catcode `\\12\catcode `\$12\catcode
  `\&12\catcode `\#12\catcode `\^12\catcode `\_12\catcode `\%12\relax}%
\providecommand \@@startlink[1]{}%
\providecommand \@@endlink[0]{}%
\providecommand \url  [0]{\begingroup\@sanitize@url \@url }%
\providecommand \@url [1]{\endgroup\@href {#1}{\urlprefix }}%
\providecommand \urlprefix  [0]{URL }%
\providecommand \Eprint [0]{\href }%
\providecommand \doibase [0]{http://dx.doi.org/}%
\providecommand \selectlanguage [0]{\@gobble}%
\providecommand \bibinfo  [0]{\@secondoftwo}%
\providecommand \bibfield  [0]{\@secondoftwo}%
\providecommand \translation [1]{[#1]}%
\providecommand \BibitemOpen [0]{}%
\providecommand \bibitemStop [0]{}%
\providecommand \bibitemNoStop [0]{.\EOS\space}%
\providecommand \EOS [0]{\spacefactor3000\relax}%
\providecommand \BibitemShut  [1]{\csname bibitem#1\endcsname}%
\let\auto@bib@innerbib\@empty
%</preamble>
\bibitem [{\citenamefont {Omidvar}\ \emph {et~al.}(2020)\citenamefont
  {Omidvar}, \citenamefont {Bou-Zeid}, \citenamefont {Li}, \citenamefont
  {Mellado},\ and\ \citenamefont {Klein}}]{omidvar2020plume}%
  \BibitemOpen
  \bibfield  {author} {\bibinfo {author} {\bibfnamefont {H.}~\bibnamefont
  {Omidvar}}, \bibinfo {author} {\bibfnamefont {E.}~\bibnamefont {Bou-Zeid}},
  \bibinfo {author} {\bibfnamefont {Q.}~\bibnamefont {Li}}, \bibinfo {author}
  {\bibfnamefont {J.-P.}\ \bibnamefont {Mellado}}, \ and\ \bibinfo {author}
  {\bibfnamefont {P.}~\bibnamefont {Klein}},\ }\bibfield  {title} {\enquote
  {\bibinfo {title} {Plume or bubble? {M}ixed-convection flow regimes and
  city-scale circulations},}\ }\href@noop {} {\bibfield  {journal} {\bibinfo
  {journal} {Journal of Fluid Mechanics}\ }\textbf {\bibinfo {volume} {897}},\
  \bibinfo {pages} {A5} (\bibinfo {year} {2020})}\BibitemShut {NoStop}%
\bibitem [{\citenamefont {Bhagat}\ \emph {et~al.}(2020)\citenamefont {Bhagat},
  \citenamefont {Wykes}, \citenamefont {Dalziel},\ and\ \citenamefont
  {Linden}}]{bhagat2020effects}%
  \BibitemOpen
  \bibfield  {author} {\bibinfo {author} {\bibfnamefont {R.~K.}\ \bibnamefont
  {Bhagat}}, \bibinfo {author} {\bibfnamefont {M.~D.}\ \bibnamefont {Wykes}},
  \bibinfo {author} {\bibfnamefont {S.~B.}\ \bibnamefont {Dalziel}}, \ and\
  \bibinfo {author} {\bibfnamefont {P.}~\bibnamefont {Linden}},\ }\bibfield
  {title} {\enquote {\bibinfo {title} {Effects of ventilation on the indoor
  spread of {COVID}-19},}\ }\href@noop {} {\bibfield  {journal} {\bibinfo
  {journal} {Journal of Fluid Mechanics}\ }\textbf {\bibinfo {volume} {903}},\
  \bibinfo {pages} {F1} (\bibinfo {year} {2020})}\BibitemShut {NoStop}%
\bibitem [{\citenamefont {Xu}\ \emph {et~al.}(2020{\natexlab{a}})\citenamefont
  {Xu}, \citenamefont {Tao}, \citenamefont {Shi},\ and\ \citenamefont
  {Xi}}]{xu2020transport}%
  \BibitemOpen
  \bibfield  {author} {\bibinfo {author} {\bibfnamefont {A.}~\bibnamefont
  {Xu}}, \bibinfo {author} {\bibfnamefont {S.}~\bibnamefont {Tao}}, \bibinfo
  {author} {\bibfnamefont {L.}~\bibnamefont {Shi}}, \ and\ \bibinfo {author}
  {\bibfnamefont {H.-D.}\ \bibnamefont {Xi}},\ }\bibfield  {title} {\enquote
  {\bibinfo {title} {Transport and deposition of dilute microparticles in
  turbulent thermal convection},}\ }\href@noop {} {\bibfield  {journal}
  {\bibinfo  {journal} {Physics of Fluids}\ }\textbf {\bibinfo {volume} {32}},\
  \bibinfo {pages} {083301} (\bibinfo {year} {2020}{\natexlab{a}})}\BibitemShut
  {NoStop}%
\bibitem [{\citenamefont {Shah}\ and\ \citenamefont
  {Sekulic}(2003)}]{shah2003fundamentals}%
  \BibitemOpen
  \bibfield  {author} {\bibinfo {author} {\bibfnamefont {R.~K.}\ \bibnamefont
  {Shah}}\ and\ \bibinfo {author} {\bibfnamefont {D.~P.}\ \bibnamefont
  {Sekulic}},\ }\href@noop {} {\emph {\bibinfo {title} {Fundamentals of heat
  exchanger design}}}\ (\bibinfo  {publisher} {John Wiley \& Sons},\ \bibinfo
  {year} {2003})\BibitemShut {NoStop}%
\bibitem [{\citenamefont {Lohse}\ and\ \citenamefont
  {Xia}(2010)}]{lohse2010small}%
  \BibitemOpen
  \bibfield  {author} {\bibinfo {author} {\bibfnamefont {D.}~\bibnamefont
  {Lohse}}\ and\ \bibinfo {author} {\bibfnamefont {K.-Q.}\ \bibnamefont
  {Xia}},\ }\bibfield  {title} {\enquote {\bibinfo {title} {Small-scale
  properties of turbulent {R}ayleigh-{B}{\'e}nard convection},}\ }\href@noop {}
  {\bibfield  {journal} {\bibinfo  {journal} {Annual Review of Fluid
  Mechanics}\ }\textbf {\bibinfo {volume} {42}},\ \bibinfo {pages} {335--364}
  (\bibinfo {year} {2010})}\BibitemShut {NoStop}%
\bibitem [{\citenamefont {Xia}(2013)}]{xia2013current}%
  \BibitemOpen
  \bibfield  {author} {\bibinfo {author} {\bibfnamefont {K.-Q.}\ \bibnamefont
  {Xia}},\ }\bibfield  {title} {\enquote {\bibinfo {title} {Current trends and
  future directions in turbulent thermal convection},}\ }\href@noop {}
  {\bibfield  {journal} {\bibinfo  {journal} {Theoretical and Applied Mechanics
  Letters}\ }\textbf {\bibinfo {volume} {3}},\ \bibinfo {pages} {052001}
  (\bibinfo {year} {2013})}\BibitemShut {NoStop}%
\bibitem [{\citenamefont {Zhang}\ \emph {et~al.}(2018)\citenamefont {Zhang},
  \citenamefont {Sun}, \citenamefont {Bao},\ and\ \citenamefont
  {Zhou}}]{zhang2018surface}%
  \BibitemOpen
  \bibfield  {author} {\bibinfo {author} {\bibfnamefont {Y.-Z.}\ \bibnamefont
  {Zhang}}, \bibinfo {author} {\bibfnamefont {C.}~\bibnamefont {Sun}}, \bibinfo
  {author} {\bibfnamefont {Y.}~\bibnamefont {Bao}}, \ and\ \bibinfo {author}
  {\bibfnamefont {Q.}~\bibnamefont {Zhou}},\ }\bibfield  {title} {\enquote
  {\bibinfo {title} {How surface roughness reduces heat transport for small
  roughness heights in turbulent {R}ayleigh--{B}{\'e}nard convection},}\
  }\href@noop {} {\bibfield  {journal} {\bibinfo  {journal} {Journal of Fluid
  Mechanics}\ }\textbf {\bibinfo {volume} {836}},\ \bibinfo {pages} {R2}
  (\bibinfo {year} {2018})}\BibitemShut {NoStop}%
\bibitem [{\citenamefont {Dong}\ \emph {et~al.}(2020)\citenamefont {Dong},
  \citenamefont {Wang}, \citenamefont {Dong}, \citenamefont {Huang},
  \citenamefont {Jiang}, \citenamefont {Liu}, \citenamefont {Lu}, \citenamefont
  {Qiu}, \citenamefont {Tang},\ and\ \citenamefont {Zhou}}]{dong2020influence}%
  \BibitemOpen
  \bibfield  {author} {\bibinfo {author} {\bibfnamefont {D.-L.}\ \bibnamefont
  {Dong}}, \bibinfo {author} {\bibfnamefont {B.-F.}\ \bibnamefont {Wang}},
  \bibinfo {author} {\bibfnamefont {Y.-H.}\ \bibnamefont {Dong}}, \bibinfo
  {author} {\bibfnamefont {Y.-X.}\ \bibnamefont {Huang}}, \bibinfo {author}
  {\bibfnamefont {N.}~\bibnamefont {Jiang}}, \bibinfo {author} {\bibfnamefont
  {Y.-L.}\ \bibnamefont {Liu}}, \bibinfo {author} {\bibfnamefont {Z.-M.}\
  \bibnamefont {Lu}}, \bibinfo {author} {\bibfnamefont {X.}~\bibnamefont
  {Qiu}}, \bibinfo {author} {\bibfnamefont {Z.-Q.}\ \bibnamefont {Tang}}, \
  and\ \bibinfo {author} {\bibfnamefont {Q.}~\bibnamefont {Zhou}},\ }\bibfield
  {title} {\enquote {\bibinfo {title} {Influence of spatial arrangements of
  roughness elements on turbulent {R}ayleigh-{B}{\'e}nard convection},}\
  }\href@noop {} {\bibfield  {journal} {\bibinfo  {journal} {Physics of
  Fluids}\ }\textbf {\bibinfo {volume} {32}},\ \bibinfo {pages} {045114}
  (\bibinfo {year} {2020})}\BibitemShut {NoStop}%
\bibitem [{\citenamefont {Huang}\ \emph {et~al.}(2013)\citenamefont {Huang},
  \citenamefont {Kaczorowski}, \citenamefont {Ni},\ and\ \citenamefont
  {Xia}}]{huang2013confinement}%
  \BibitemOpen
  \bibfield  {author} {\bibinfo {author} {\bibfnamefont {S.-D.}\ \bibnamefont
  {Huang}}, \bibinfo {author} {\bibfnamefont {M.}~\bibnamefont {Kaczorowski}},
  \bibinfo {author} {\bibfnamefont {R.}~\bibnamefont {Ni}}, \ and\ \bibinfo
  {author} {\bibfnamefont {K.-Q.}\ \bibnamefont {Xia}},\ }\bibfield  {title}
  {\enquote {\bibinfo {title} {Confinement-induced heat-transport enhancement
  in turbulent thermal convection},}\ }\href@noop {} {\bibfield  {journal}
  {\bibinfo  {journal} {Physical Review Letters}\ }\textbf {\bibinfo {volume}
  {111}},\ \bibinfo {pages} {104501} (\bibinfo {year} {2013})}\BibitemShut
  {NoStop}%
\bibitem [{\citenamefont {Huang}\ and\ \citenamefont
  {Xia}(2016)}]{huang2016effects}%
  \BibitemOpen
  \bibfield  {author} {\bibinfo {author} {\bibfnamefont {S.-D.}\ \bibnamefont
  {Huang}}\ and\ \bibinfo {author} {\bibfnamefont {K.-Q.}\ \bibnamefont
  {Xia}},\ }\bibfield  {title} {\enquote {\bibinfo {title} {Effects of
  geometric confinement in quasi-2-{D} turbulent {R}ayleigh--{B}{\'e}nard
  convection},}\ }\href@noop {} {\bibfield  {journal} {\bibinfo  {journal}
  {Journal of Fluid Mechanics}\ }\textbf {\bibinfo {volume} {794}},\ \bibinfo
  {pages} {639--654} (\bibinfo {year} {2016})}\BibitemShut {NoStop}%
\bibitem [{\citenamefont {Wang}, \citenamefont {Zhou},\ and\ \citenamefont
  {Sun}(2020)}]{wang2020vibration}%
  \BibitemOpen
  \bibfield  {author} {\bibinfo {author} {\bibfnamefont {B.-F.}\ \bibnamefont
  {Wang}}, \bibinfo {author} {\bibfnamefont {Q.}~\bibnamefont {Zhou}}, \ and\
  \bibinfo {author} {\bibfnamefont {C.}~\bibnamefont {Sun}},\ }\bibfield
  {title} {\enquote {\bibinfo {title} {Vibration-induced boundary-layer
  destabilization achieves massive heat-transport enhancement},}\ }\href@noop
  {} {\bibfield  {journal} {\bibinfo  {journal} {Science Advances}\ }\textbf
  {\bibinfo {volume} {6}},\ \bibinfo {pages} {eaaz8239} (\bibinfo {year}
  {2020})}\BibitemShut {NoStop}%
\bibitem [{\citenamefont {Wu}\ \emph {et~al.}(2021)\citenamefont {Wu},
  \citenamefont {Dong}, \citenamefont {Wang},\ and\ \citenamefont
  {Zhou}}]{wu2021phase}%
  \BibitemOpen
  \bibfield  {author} {\bibinfo {author} {\bibfnamefont {J.-Z.}\ \bibnamefont
  {Wu}}, \bibinfo {author} {\bibfnamefont {Y.-H.}\ \bibnamefont {Dong}},
  \bibinfo {author} {\bibfnamefont {B.-F.}\ \bibnamefont {Wang}}, \ and\
  \bibinfo {author} {\bibfnamefont {Q.}~\bibnamefont {Zhou}},\ }\bibfield
  {title} {\enquote {\bibinfo {title} {Phase decomposition analysis on
  oscillatory {R}ayleigh--{B}{\'e}nard turbulence},}\ }\href@noop {} {\bibfield
   {journal} {\bibinfo  {journal} {Physics of Fluids}\ }\textbf {\bibinfo
  {volume} {33}},\ \bibinfo {pages} {045108} (\bibinfo {year}
  {2021})}\BibitemShut {NoStop}%
\bibitem [{\citenamefont {Sun}, \citenamefont {Xi},\ and\ \citenamefont
  {Xia}(2005)}]{sun2005azimuthal}%
  \BibitemOpen
  \bibfield  {author} {\bibinfo {author} {\bibfnamefont {C.}~\bibnamefont
  {Sun}}, \bibinfo {author} {\bibfnamefont {H.-D.}\ \bibnamefont {Xi}}, \ and\
  \bibinfo {author} {\bibfnamefont {K.-Q.}\ \bibnamefont {Xia}},\ }\bibfield
  {title} {\enquote {\bibinfo {title} {Azimuthal symmetry, flow dynamics, and
  heat transport in turbulent thermal convection in a cylinder with an aspect
  ratio of 0.5},}\ }\href@noop {} {\bibfield  {journal} {\bibinfo  {journal}
  {Physical Review Letters}\ }\textbf {\bibinfo {volume} {95}},\ \bibinfo
  {pages} {074502} (\bibinfo {year} {2005})}\BibitemShut {NoStop}%
\bibitem [{\citenamefont {Xi}\ and\ \citenamefont {Xia}(2008)}]{xi2008flow}%
  \BibitemOpen
  \bibfield  {author} {\bibinfo {author} {\bibfnamefont {H.-D.}\ \bibnamefont
  {Xi}}\ and\ \bibinfo {author} {\bibfnamefont {K.-Q.}\ \bibnamefont {Xia}},\
  }\bibfield  {title} {\enquote {\bibinfo {title} {Flow mode transitions in
  turbulent thermal convection},}\ }\href@noop {} {\bibfield  {journal}
  {\bibinfo  {journal} {Physics of Fluids}\ }\textbf {\bibinfo {volume} {20}},\
  \bibinfo {pages} {055104} (\bibinfo {year} {2008})}\BibitemShut {NoStop}%
\bibitem [{\citenamefont {Xi}\ \emph {et~al.}(2016)\citenamefont {Xi},
  \citenamefont {Zhang}, \citenamefont {Hao},\ and\ \citenamefont
  {Xia}}]{xi2016higher}%
  \BibitemOpen
  \bibfield  {author} {\bibinfo {author} {\bibfnamefont {H.-D.}\ \bibnamefont
  {Xi}}, \bibinfo {author} {\bibfnamefont {Y.-B.}\ \bibnamefont {Zhang}},
  \bibinfo {author} {\bibfnamefont {J.-T.}\ \bibnamefont {Hao}}, \ and\
  \bibinfo {author} {\bibfnamefont {K.-Q.}\ \bibnamefont {Xia}},\ }\bibfield
  {title} {\enquote {\bibinfo {title} {Higher-order flow modes in turbulent
  {R}ayleigh--{B}{\'e}nard convection},}\ }\href@noop {} {\bibfield  {journal}
  {\bibinfo  {journal} {Journal of Fluid Mechanics}\ }\textbf {\bibinfo
  {volume} {805}},\ \bibinfo {pages} {31--51} (\bibinfo {year}
  {2016})}\BibitemShut {NoStop}%
\bibitem [{\citenamefont {Xu}\ \emph {et~al.}(2020{\natexlab{b}})\citenamefont
  {Xu}, \citenamefont {Chen}, \citenamefont {Wang},\ and\ \citenamefont
  {Xi}}]{xu2020correlation}%
  \BibitemOpen
  \bibfield  {author} {\bibinfo {author} {\bibfnamefont {A.}~\bibnamefont
  {Xu}}, \bibinfo {author} {\bibfnamefont {X.}~\bibnamefont {Chen}}, \bibinfo
  {author} {\bibfnamefont {F.}~\bibnamefont {Wang}}, \ and\ \bibinfo {author}
  {\bibfnamefont {H.-D.}\ \bibnamefont {Xi}},\ }\bibfield  {title} {\enquote
  {\bibinfo {title} {Correlation of internal flow structure with heat transfer
  efficiency in turbulent {R}ayleigh--{B}{\'e}nard convection},}\ }\href@noop
  {} {\bibfield  {journal} {\bibinfo  {journal} {Physics of Fluids}\ }\textbf
  {\bibinfo {volume} {32}},\ \bibinfo {pages} {105112} (\bibinfo {year}
  {2020}{\natexlab{b}})}\BibitemShut {NoStop}%
\bibitem [{\citenamefont {van~der Poel}, \citenamefont {Stevens},\ and\
  \citenamefont {Lohse}(2011)}]{van2011connecting}%
  \BibitemOpen
  \bibfield  {author} {\bibinfo {author} {\bibfnamefont {E.~P.}\ \bibnamefont
  {van~der Poel}}, \bibinfo {author} {\bibfnamefont {R.~J.}\ \bibnamefont
  {Stevens}}, \ and\ \bibinfo {author} {\bibfnamefont {D.}~\bibnamefont
  {Lohse}},\ }\bibfield  {title} {\enquote {\bibinfo {title} {Connecting flow
  structures and heat flux in turbulent {R}ayleigh-{B}{\'e}nard convection},}\
  }\href@noop {} {\bibfield  {journal} {\bibinfo  {journal} {Physical Review
  E}\ }\textbf {\bibinfo {volume} {84}},\ \bibinfo {pages} {045303} (\bibinfo
  {year} {2011})}\BibitemShut {NoStop}%
\bibitem [{\citenamefont {van~der Poel}\ \emph {et~al.}(2012)\citenamefont
  {van~der Poel}, \citenamefont {Stevens}, \citenamefont {Sugiyama},\ and\
  \citenamefont {Lohse}}]{van2012flow}%
  \BibitemOpen
  \bibfield  {author} {\bibinfo {author} {\bibfnamefont {E.~P.}\ \bibnamefont
  {van~der Poel}}, \bibinfo {author} {\bibfnamefont {R.~J.}\ \bibnamefont
  {Stevens}}, \bibinfo {author} {\bibfnamefont {K.}~\bibnamefont {Sugiyama}}, \
  and\ \bibinfo {author} {\bibfnamefont {D.}~\bibnamefont {Lohse}},\ }\bibfield
   {title} {\enquote {\bibinfo {title} {Flow states in two-dimensional
  {R}ayleigh-{B}{\'e}nard convection as a function of aspect-ratio and
  {R}ayleigh number},}\ }\href@noop {} {\bibfield  {journal} {\bibinfo
  {journal} {Physics of Fluids}\ }\textbf {\bibinfo {volume} {24}},\ \bibinfo
  {pages} {085104} (\bibinfo {year} {2012})}\BibitemShut {NoStop}%
\bibitem [{\citenamefont {Zou}\ \emph {et~al.}(2019)\citenamefont {Zou},
  \citenamefont {Zhou}, \citenamefont {Chen}, \citenamefont {Bao},
  \citenamefont {Chen},\ and\ \citenamefont {She}}]{zou2019boundary}%
  \BibitemOpen
  \bibfield  {author} {\bibinfo {author} {\bibfnamefont {H.-Y.}\ \bibnamefont
  {Zou}}, \bibinfo {author} {\bibfnamefont {W.-F.}\ \bibnamefont {Zhou}},
  \bibinfo {author} {\bibfnamefont {X.}~\bibnamefont {Chen}}, \bibinfo {author}
  {\bibfnamefont {Y.}~\bibnamefont {Bao}}, \bibinfo {author} {\bibfnamefont
  {J.}~\bibnamefont {Chen}}, \ and\ \bibinfo {author} {\bibfnamefont {Z.-S.}\
  \bibnamefont {She}},\ }\bibfield  {title} {\enquote {\bibinfo {title}
  {Boundary layer structure in turbulent {R}ayleigh--{B}{\'e}nard convection in
  a slim box},}\ }\href@noop {} {\bibfield  {journal} {\bibinfo  {journal}
  {Acta Mechanica Sinica}\ }\textbf {\bibinfo {volume} {35}},\ \bibinfo {pages}
  {713--728} (\bibinfo {year} {2019})}\BibitemShut {NoStop}%
\bibitem [{\citenamefont {Zhu}\ and\ \citenamefont {Zhou}(2021)}]{zhu2021flow}%
  \BibitemOpen
  \bibfield  {author} {\bibinfo {author} {\bibfnamefont {X.}~\bibnamefont
  {Zhu}}\ and\ \bibinfo {author} {\bibfnamefont {Q.}~\bibnamefont {Zhou}},\
  }\bibfield  {title} {\enquote {\bibinfo {title} {Flow structures of turbulent
  {R}ayleigh--{B}{\'e}nard convection in annular cells with aspect ratio one
  and larger},}\ }\href@noop {} {\bibfield  {journal} {\bibinfo  {journal}
  {Acta Mechanica Sinica}\ }\textbf {\bibinfo {volume} {37}},\ \bibinfo {pages}
  {1291--1298} (\bibinfo {year} {2021})}\BibitemShut {NoStop}%
\bibitem [{\citenamefont {Panton}(2013)}]{panton2013incompressible}%
  \BibitemOpen
  \bibfield  {author} {\bibinfo {author} {\bibfnamefont {R.~L.}\ \bibnamefont
  {Panton}},\ }\href@noop {} {\emph {\bibinfo {title} {Incompressible flow}}}\
  (\bibinfo  {publisher} {John Wiley \& Sons},\ \bibinfo {year}
  {2013})\BibitemShut {NoStop}%
\bibitem [{\citenamefont {Davidson}(2015)}]{davidson2015turbulence}%
  \BibitemOpen
  \bibfield  {author} {\bibinfo {author} {\bibfnamefont {P.~A.}\ \bibnamefont
  {Davidson}},\ }\href@noop {} {\emph {\bibinfo {title} {Turbulence: {A}n
  introduction for scientists and engineers}}}\ (\bibinfo  {publisher} {Oxford
  university press},\ \bibinfo {year} {2015})\BibitemShut {NoStop}%
\bibitem [{\citenamefont {Lemenand}\ \emph {et~al.}(2018)\citenamefont
  {Lemenand}, \citenamefont {Habchi}, \citenamefont {Della~Valle},\ and\
  \citenamefont {Peerhossaini}}]{lemenand2018vorticity}%
  \BibitemOpen
  \bibfield  {author} {\bibinfo {author} {\bibfnamefont {T.}~\bibnamefont
  {Lemenand}}, \bibinfo {author} {\bibfnamefont {C.}~\bibnamefont {Habchi}},
  \bibinfo {author} {\bibfnamefont {D.}~\bibnamefont {Della~Valle}}, \ and\
  \bibinfo {author} {\bibfnamefont {H.}~\bibnamefont {Peerhossaini}},\
  }\bibfield  {title} {\enquote {\bibinfo {title} {Vorticity and convective
  heat transfer downstream of a vortex generator},}\ }\href@noop {} {\bibfield
  {journal} {\bibinfo  {journal} {International Journal of Thermal Sciences}\
  }\textbf {\bibinfo {volume} {125}},\ \bibinfo {pages} {342--349} (\bibinfo
  {year} {2018})}\BibitemShut {NoStop}%
\bibitem [{\citenamefont {Karkaba}\ \emph {et~al.}(2020)\citenamefont
  {Karkaba}, \citenamefont {Dbouk}, \citenamefont {Habchi}, \citenamefont
  {Russeil}, \citenamefont {Lemenand},\ and\ \citenamefont
  {Bougeard}}]{karkaba2020multi}%
  \BibitemOpen
  \bibfield  {author} {\bibinfo {author} {\bibfnamefont {H.}~\bibnamefont
  {Karkaba}}, \bibinfo {author} {\bibfnamefont {T.}~\bibnamefont {Dbouk}},
  \bibinfo {author} {\bibfnamefont {C.}~\bibnamefont {Habchi}}, \bibinfo
  {author} {\bibfnamefont {S.}~\bibnamefont {Russeil}}, \bibinfo {author}
  {\bibfnamefont {T.}~\bibnamefont {Lemenand}}, \ and\ \bibinfo {author}
  {\bibfnamefont {D.}~\bibnamefont {Bougeard}},\ }\bibfield  {title} {\enquote
  {\bibinfo {title} {Multi objective optimization of vortex generators for heat
  transfer enhancement using large design space exploration},}\ }\href@noop {}
  {\bibfield  {journal} {\bibinfo  {journal} {Chemical Engineering and
  Processing-Process Intensification}\ }\textbf {\bibinfo {volume} {154}},\
  \bibinfo {pages} {107982} (\bibinfo {year} {2020})}\BibitemShut {NoStop}%
\bibitem [{\citenamefont {Chen}\ and\ \citenamefont
  {Doolen}(1998)}]{chen1998lattice}%
  \BibitemOpen
  \bibfield  {author} {\bibinfo {author} {\bibfnamefont {S.}~\bibnamefont
  {Chen}}\ and\ \bibinfo {author} {\bibfnamefont {G.~D.}\ \bibnamefont
  {Doolen}},\ }\bibfield  {title} {\enquote {\bibinfo {title} {Lattice
  {B}oltzmann method for fluid flows},}\ }\href@noop {} {\bibfield  {journal}
  {\bibinfo  {journal} {Annual Review of Fluid Mechanics}\ }\textbf {\bibinfo
  {volume} {30}},\ \bibinfo {pages} {329--364} (\bibinfo {year}
  {1998})}\BibitemShut {NoStop}%
\bibitem [{\citenamefont {Aidun}\ and\ \citenamefont
  {Clausen}(2010)}]{aidun2010lattice}%
  \BibitemOpen
  \bibfield  {author} {\bibinfo {author} {\bibfnamefont {C.~K.}\ \bibnamefont
  {Aidun}}\ and\ \bibinfo {author} {\bibfnamefont {J.~R.}\ \bibnamefont
  {Clausen}},\ }\bibfield  {title} {\enquote {\bibinfo {title}
  {Lattice-{B}oltzmann method for complex flows},}\ }\href@noop {} {\bibfield
  {journal} {\bibinfo  {journal} {Annual Review of Fluid Mechanics}\ }\textbf
  {\bibinfo {volume} {42}},\ \bibinfo {pages} {439--472} (\bibinfo {year}
  {2010})}\BibitemShut {NoStop}%
\bibitem [{\citenamefont {Xu}, \citenamefont {Shyy},\ and\ \citenamefont
  {Zhao}(2017)}]{xu2017lattice}%
  \BibitemOpen
  \bibfield  {author} {\bibinfo {author} {\bibfnamefont {A.}~\bibnamefont
  {Xu}}, \bibinfo {author} {\bibfnamefont {W.}~\bibnamefont {Shyy}}, \ and\
  \bibinfo {author} {\bibfnamefont {T.}~\bibnamefont {Zhao}},\ }\bibfield
  {title} {\enquote {\bibinfo {title} {Lattice {B}oltzmann modeling of
  transport phenomena in fuel cells and flow batteries},}\ }\href@noop {}
  {\bibfield  {journal} {\bibinfo  {journal} {Acta Mechanica Sinica}\ }\textbf
  {\bibinfo {volume} {33}},\ \bibinfo {pages} {555--574} (\bibinfo {year}
  {2017})}\BibitemShut {NoStop}%
\bibitem [{\citenamefont {Xu}, \citenamefont {Shi},\ and\ \citenamefont
  {Zhao}(2017)}]{xu2017accelerated}%
  \BibitemOpen
  \bibfield  {author} {\bibinfo {author} {\bibfnamefont {A.}~\bibnamefont
  {Xu}}, \bibinfo {author} {\bibfnamefont {L.}~\bibnamefont {Shi}}, \ and\
  \bibinfo {author} {\bibfnamefont {T.}~\bibnamefont {Zhao}},\ }\bibfield
  {title} {\enquote {\bibinfo {title} {Accelerated lattice {B}oltzmann
  simulation using {GPU} and {O}pen{ACC} with data management},}\ }\href@noop
  {} {\bibfield  {journal} {\bibinfo  {journal} {International Journal of Heat
  and Mass Transfer}\ }\textbf {\bibinfo {volume} {109}},\ \bibinfo {pages}
  {577--588} (\bibinfo {year} {2017})}\BibitemShut {NoStop}%
\bibitem [{\citenamefont {Xu}, \citenamefont {Shi},\ and\ \citenamefont
  {Xi}(2019{\natexlab{a}})}]{xu2019lattice}%
  \BibitemOpen
  \bibfield  {author} {\bibinfo {author} {\bibfnamefont {A.}~\bibnamefont
  {Xu}}, \bibinfo {author} {\bibfnamefont {L.}~\bibnamefont {Shi}}, \ and\
  \bibinfo {author} {\bibfnamefont {H.-D.}\ \bibnamefont {Xi}},\ }\bibfield
  {title} {\enquote {\bibinfo {title} {Lattice {B}oltzmann simulations of
  three-dimensional thermal convective flows at high {R}ayleigh number},}\
  }\href@noop {} {\bibfield  {journal} {\bibinfo  {journal} {International
  Journal of Heat and Mass Transfer}\ }\textbf {\bibinfo {volume} {140}},\
  \bibinfo {pages} {359--370} (\bibinfo {year}
  {2019}{\natexlab{a}})}\BibitemShut {NoStop}%
\bibitem [{\citenamefont {Xu}, \citenamefont {Shi},\ and\ \citenamefont
  {Xi}(2019{\natexlab{b}})}]{xu2019statistics}%
  \BibitemOpen
  \bibfield  {author} {\bibinfo {author} {\bibfnamefont {A.}~\bibnamefont
  {Xu}}, \bibinfo {author} {\bibfnamefont {L.}~\bibnamefont {Shi}}, \ and\
  \bibinfo {author} {\bibfnamefont {H.-D.}\ \bibnamefont {Xi}},\ }\bibfield
  {title} {\enquote {\bibinfo {title} {Statistics of temperature and thermal
  energy dissipation rate in low-{P}randtl number turbulent thermal
  convection},}\ }\href@noop {} {\bibfield  {journal} {\bibinfo  {journal}
  {Physics of Fluids}\ }\textbf {\bibinfo {volume} {31}},\ \bibinfo {pages}
  {125101} (\bibinfo {year} {2019}{\natexlab{b}})}\BibitemShut {NoStop}%
\bibitem [{\citenamefont {Xu}, \citenamefont {Chen},\ and\ \citenamefont
  {Xi}(2021)}]{xu2021tristable}%
  \BibitemOpen
  \bibfield  {author} {\bibinfo {author} {\bibfnamefont {A.}~\bibnamefont
  {Xu}}, \bibinfo {author} {\bibfnamefont {X.}~\bibnamefont {Chen}}, \ and\
  \bibinfo {author} {\bibfnamefont {H.-D.}\ \bibnamefont {Xi}},\ }\bibfield
  {title} {\enquote {\bibinfo {title} {Tristable flow states and reversal of
  the large-scale circulation in two-dimensional circular convection cells},}\
  }\href@noop {} {\bibfield  {journal} {\bibinfo  {journal} {Journal of Fluid
  Mechanics}\ }\textbf {\bibinfo {volume} {910}},\ \bibinfo {pages} {A33}
  (\bibinfo {year} {2021})}\BibitemShut {NoStop}%
\bibitem [{\citenamefont {Kenjere{\v{s}}}\ and\ \citenamefont
  {Hanjali{\'c}}(2000)}]{kenjerevs2000convective}%
  \BibitemOpen
  \bibfield  {author} {\bibinfo {author} {\bibfnamefont {S.}~\bibnamefont
  {Kenjere{\v{s}}}}\ and\ \bibinfo {author} {\bibfnamefont {K.}~\bibnamefont
  {Hanjali{\'c}}},\ }\bibfield  {title} {\enquote {\bibinfo {title} {Convective
  rolls and heat transfer in finite-length {R}ayleigh-{B}{\'e}nard convection:
  {A} two-dimensional numerical study},}\ }\href@noop {} {\bibfield  {journal}
  {\bibinfo  {journal} {Physical Review E}\ }\textbf {\bibinfo {volume} {62}},\
  \bibinfo {pages} {7987} (\bibinfo {year} {2000})}\BibitemShut {NoStop}%
\bibitem [{\citenamefont {Wu}, \citenamefont {Ma},\ and\ \citenamefont
  {Zhou}(2007)}]{wu2007vorticity}%
  \BibitemOpen
  \bibfield  {author} {\bibinfo {author} {\bibfnamefont {J.-Z.}\ \bibnamefont
  {Wu}}, \bibinfo {author} {\bibfnamefont {H.-Y.}\ \bibnamefont {Ma}}, \ and\
  \bibinfo {author} {\bibfnamefont {M.-D.}\ \bibnamefont {Zhou}},\ }\href@noop
  {} {\emph {\bibinfo {title} {Vorticity and vortex dynamics}}}\ (\bibinfo
  {publisher} {Springer},\ \bibinfo {year} {2007})\BibitemShut {NoStop}%
\bibitem [{\citenamefont {Hunt}, \citenamefont {Wray},\ and\ \citenamefont
  {Moin}(1988)}]{jcr1988eddies}%
  \BibitemOpen
  \bibfield  {author} {\bibinfo {author} {\bibfnamefont {J.}~\bibnamefont
  {Hunt}}, \bibinfo {author} {\bibfnamefont {A.}~\bibnamefont {Wray}}, \ and\
  \bibinfo {author} {\bibfnamefont {P.}~\bibnamefont {Moin}},\ }\bibfield
  {title} {\enquote {\bibinfo {title} {Eddies, stream, and convergence zones in
  turbulent flows},}\ }\href@noop {} {\bibfield  {journal} {\bibinfo  {journal}
  {Center for Turbulence Research Report}\ }\textbf {\bibinfo {volume}
  {CTR-S88}},\ \bibinfo {pages} {193} (\bibinfo {year} {1988})}\BibitemShut
  {NoStop}%
\bibitem [{\citenamefont {Jeong}\ and\ \citenamefont
  {Hussain}(1995)}]{jeong1995identification}%
  \BibitemOpen
  \bibfield  {author} {\bibinfo {author} {\bibfnamefont {J.}~\bibnamefont
  {Jeong}}\ and\ \bibinfo {author} {\bibfnamefont {F.}~\bibnamefont
  {Hussain}},\ }\bibfield  {title} {\enquote {\bibinfo {title} {On the
  identification of a vortex},}\ }\href@noop {} {\bibfield  {journal} {\bibinfo
   {journal} {Journal of Fluid Mechanics}\ }\textbf {\bibinfo {volume} {285}},\
  \bibinfo {pages} {69--94} (\bibinfo {year} {1995})}\BibitemShut {NoStop}%
\bibitem [{\citenamefont {Chong}, \citenamefont {Perry},\ and\ \citenamefont
  {Cantwell}(1990)}]{chong1990general}%
  \BibitemOpen
  \bibfield  {author} {\bibinfo {author} {\bibfnamefont {M.~S.}\ \bibnamefont
  {Chong}}, \bibinfo {author} {\bibfnamefont {A.~E.}\ \bibnamefont {Perry}}, \
  and\ \bibinfo {author} {\bibfnamefont {B.~J.}\ \bibnamefont {Cantwell}},\
  }\bibfield  {title} {\enquote {\bibinfo {title} {A general classification of
  three-dimensional flow fields},}\ }\href@noop {} {\bibfield  {journal}
  {\bibinfo  {journal} {Physics of Fluids A: Fluid Dynamics}\ }\textbf
  {\bibinfo {volume} {2}},\ \bibinfo {pages} {765--777} (\bibinfo {year}
  {1990})}\BibitemShut {NoStop}%
\bibitem [{\citenamefont {Zhou}\ \emph {et~al.}(1999)\citenamefont {Zhou},
  \citenamefont {Adrian}, \citenamefont {Balachandar},\ and\ \citenamefont
  {Kendall}}]{zhou1999mechanisms}%
  \BibitemOpen
  \bibfield  {author} {\bibinfo {author} {\bibfnamefont {J.}~\bibnamefont
  {Zhou}}, \bibinfo {author} {\bibfnamefont {R.~J.}\ \bibnamefont {Adrian}},
  \bibinfo {author} {\bibfnamefont {S.}~\bibnamefont {Balachandar}}, \ and\
  \bibinfo {author} {\bibfnamefont {T.}~\bibnamefont {Kendall}},\ }\bibfield
  {title} {\enquote {\bibinfo {title} {Mechanisms for generating coherent
  packets of hairpin vortices in channel flow},}\ }\href@noop {} {\bibfield
  {journal} {\bibinfo  {journal} {Journal of Fluid Mechanics}\ }\textbf
  {\bibinfo {volume} {387}},\ \bibinfo {pages} {353--396} (\bibinfo {year}
  {1999})}\BibitemShut {NoStop}%
\end{thebibliography}%

\end{document}